# Bivariate, Cluster and Suitability Analysis of NoSQL Solutions for Different Application Areas

Samiya Khan[a,1], Xiufeng Liu[b], Syed Arshad Ali[a], Mansaf Alam[a,2]

[a]*Jamia Millia Islamia, New Delhi, India*
[b]*Technical University of Denmark, Denmark*

## Highlights

- Performs feature analysis of 80 NoSQL solutions in view of technology selection criteria for big data systems and evaluates applications that use one of more of the mentioned NoSQL solutions.
- Presents quantitative analysis of a dataset created from qualitative analysis of 80 NoSQL solutions, investigating relationships between the chosen features.
- Proposes a classification scheme that uses 9 features, which are 'supported data model' (document-oriented, graph, key-value and wide-column), CAP (consistency, availability and partition tolerance) characteristics and free/proprietary ownership, on the basis of results obtained for k-modes clustering of the dataset.
- Determines relevant features for each class of applications and proposes a prediction model for determining the suitability of a NoSQL solution for a class of applications.

## Abstract

Big data systems' development is full of challenges in view of the variety of application areas and domains that this technology promises to serve. Typically, fundamental design decisions involved in big data systems' design include choosing appropriate storage and computing infrastructures. In this age of heterogeneous systems that integrate different technologies for development of an optimized solution to a specific real-world problem, big data systems are not an exception to any such rule. As far as the storage aspect of any big data system is concerned, the primary facet in this regard is a storage infrastructure and NoSQL is the right technology that fulfills its requirements. However, every big data application has variable data characteristics and thus, the corresponding data fits into a different data model. Moreover, the requirements of different applications vary on the basis of budget and functionality. This paper presents a feature analysis of 80 NoSQL solutions, elaborating on the criteria and points that a developer must consider while making a possible choice. Bivariate analysis of dataset created for the identified NoSQL solutions was performed to establish relationship between 9 features. Furthermore, cluster analysis of the dataset was used to create categories of solutions to present a statistically supported classification scheme. Finally, applications for different solutions were reviewed and classified under domain-specific categories. Random forest classification was used to determine the most relevant features for applications and correspondingly a decision tree-based prediction model was proposed, implemented and deployed in the form of a web application to determine the suitability of a NoSQL solution for an application area.

## 1.0 Introduction

Data is processed to generate information, which can later be used for varied purposes. Knowledge discovery and data mining are two fields that have been actively working towards deriving useful information from raw data to create applications that can make predictions, identify patterns and facilitate decision making [1]. However, with the

Corresponding authors.
E-mail addresses: [1]samiyashaukat@yahoo.com (Samiya Khan), [2]malam2@jmi.ac.in (Mansaf Alam)

rise of social media and smart devices, data is no longer a simple dataset that traditional tools and technologies can handle [2].

Digitization and rising popularity of modern technologies like smart phones and gadgets has contributed immensely towards 'data deluge'. Moreover, this data is not just high on volume, but it also includes data of varied kinds that is generated on a periodic basis. The biggest challenge in dealing with this 'big data problem' is that the present or traditional systems are unable to store and process data of this kind. Therefore, this gave rise to the need for scalable systems that can store varied forms of data and process the same to generate useful analytical solutions [3].

The present era can rightly be called the era of analytics where organizations are tapping the business potential of data by processing and analyzing it. A plethora of technologies are available for this purpose and organizations are smoothly drifting towards heterogeneous environments, which include data stores like HBase [4], HDFS [5] and MongoDB [6], execution engines like Impala [7] and Spark [8], and programming languages like R [9] and Python [10].

Big data storage [11] is a general term used for describing storage infrastructures designed for storage, management and retrieval of data that is typically high in velocity, diverse in variety and large in volume. In such infrastructures, data is stored in such a manner that its usage, processing and access become easier. Moreover, such infrastructures can scale as per the requirement of the application or service.

The primary task of big data storage is to support input and output operations on stored data in addition to storage of a large number of files and objects. Typically, the architectures used for storage of big data include a cluster of network-attached storage, pools of direct attached storage or storage based on object storage format [11]. Computing server nodes are used at the heart of these infrastructures in order to provide support for retrieval and processing of big data. Most of these storage infrastructures provide support for big data storage solutions like Hadoop [12] and NoSQL [13].

The storage needs of a big data problem are influenced by many factors. Scalability is undoubtedly one of the fundamental requirements in view of the ever-growing size of data. Any solution made for big data must be able to accommodate the growing data in an optimal manner. Considering the fact that most big data solutions require real-time analysis of data and its visualization, the allowable time in which data must be accessed is extremely low. Moreover, data needs to be accessed in a frequent and efficient manner, making availability a crucial system requirement. Big data solutions may make use of organization-specific data. One of the biggest concerns of organizations in the adoption of such solutions is the security of their data. There may be a need for big data solutions to interact with other technologies and applications. Therefore, big data solutions must be able to integrate with these technologies to create a complete application.

Technologies or solutions must be chosen on the basis of the specific requirements of a business or application. The available big data technologies offer different degrees of performance, security, data capacity and integration capabilities [14]. Therefore, if the requirements are clear and precise, choosing a solution or combination of solutions to satisfy the needs should not be difficult. This paper performs a comparative study of 80 NoSQL solutions available for use in big data systems and elaborates on how requirements must be analyzed to determine the best solution for a concerned application.

The qualitative study of NoSQL solutions was transformed into a dataset with 9 features namely, document-oriented, graph, key-value, wide-column, consistent, available, partition-tolerant, free and proprietary. It is important to mention that database type is not a single feature and the work takes document-oriented, graph, key-value and wide-column as four different features because a NoSQL solution can support multiple data models. Bivariate analysis of the dataset was performed to explore relationships between the different features. The results of this analysis indicated dependency between features and therefore, in order to propose a classification scheme, cluster analysis was performed. Applications supported by different NoSQL solutions were identified and reviewed. This analysis was further used to determine relevant features for specific classes of applications, which was in turn used to propose a prediction model for determination of suitability of a NoSQL solution for an area of application.

There are many studies on NoSQL stores. However, most of the existing literature compares the performance of a few NoSQL stores [73, 114] or provide a qualitative study of the technology [18, 19, 20, 32, 47, 81, 133]. None of the available studies in this field have elaborated upon applications of NoSQL solutions. Moreover, none of the available studies provide statistical analysis of NoSQL solutions' dataset or propose a suitability model for NoSQL applications. This paper performs exploratory analysis of the NoSQL dataset to explore the relationships between different features. Moreover, the coverage of this study, in terms of the number of NoSQL solutions used, is largest among available papers.

The rest of the paper has been organized in the manner discussed below. A brief outline of the study is presented in Section 2.0 while Section 3.0 describes NoSQL and its characteristics. It aims to answer how NoSQL solves the multiple issues presented by the big data problem to traditional systems. Classification criteria have been discussed in Section 4.0. It includes elaborate explanation on data models, CAP characteristics and other miscellaneous features of NoSQL solutions. Application domains have been discussed in Section 5.0. A dataset of 80 NoSQL solutions along with their features, which include data model, CAP characteristics and ownership, was created. Exploratory analysis of this dataset is presented in Section 6.0. A model to solve the problem of choosing a NoSQL solution for a big data system has been proposed in Section 7.0 and Section 8.0 provides an intensive discussion on technical and non-technical aspects of choosing a NoSQL solution for a big data system. Lastly, the paper concludes and summarizes the findings in Section 9.0, providing insights for future work in this field.

## 2.0 Outline of Study

This study evaluates existing NoSQL solutions from qualitative as well as quantitative perspectives. Existing literature suggests that NoSQL solutions are commonly classified on the basis of supported data model [41] and CAP characteristics [63]. Therefore, the qualitative study focuses on feature identification of 80 NoSQL solutions. These features were divided into three categories namely, supported data model, CAP characteristics and others, which included any other features that a solution supports. This analysis was used to create a dataset for NoSQL solutions and their features. The features chosen include document-oriented, graph, key-value, wide-column, consistent, available, partition-tolerant, free and proprietary. Exploratory analysis of this dataset was performed to reveal existence of relationships between features. The techniques used were Spearman's rank correlation [292] coefficient and Chi-square test [293]. Moreover, cluster analysis using k-modes clustering [297] was also performed to form groups of solutions that can further be used for classification of NoSQL solutions.

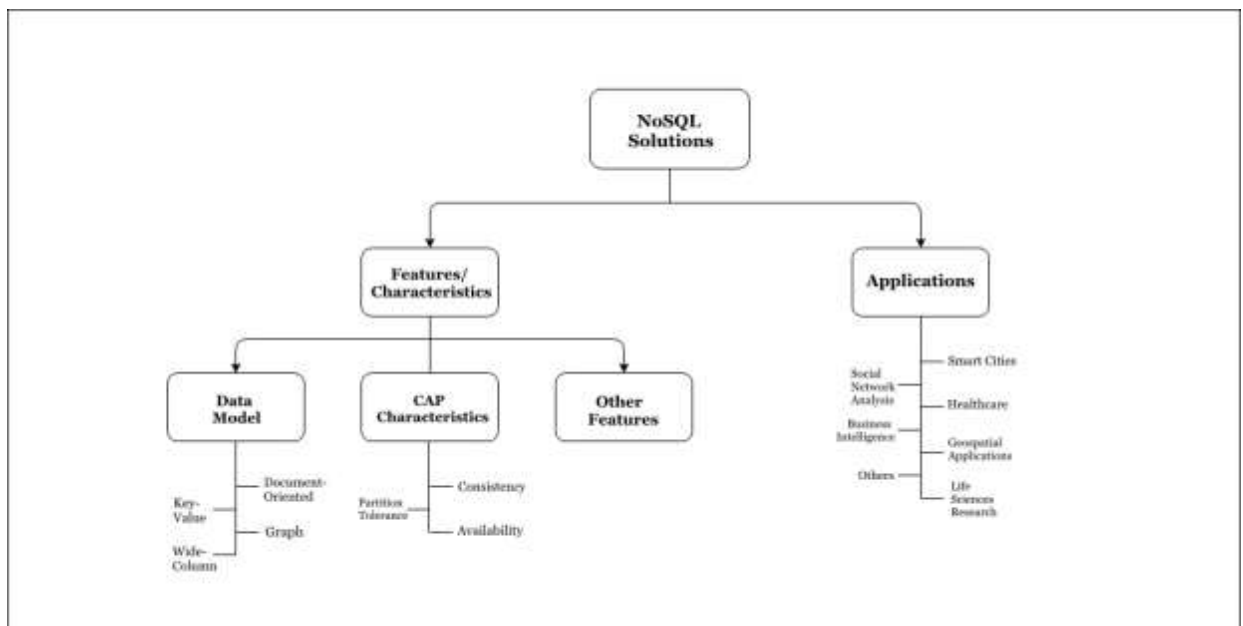

Fig. 1. Overview of Areas Covered

Besides this, applications supported by individual solutions used for analysis, were identified. On the basis of qualitative analysis of applications, they were classified under 7 categories namely, 'smart cities', 'business intelligence', 'life sciences research', 'healthcare', 'social network analysis', 'geospatial applications' and 'others'. Dataset was enhanced to incorporate supported application area. Relevant features for each application area were determined and a prediction model for determining if a NoSQL solution is suitable for an application area was proposed. The techniques used for this purpose were random forest classification [299] and decision tree classification [155]. Fig. 1 gives an overview of the areas covered in this study of NoSQL solutions.

### 3.0 NoSQL – A Solution for Big Data Storage Issues

The idea behind the development of relational databases was to provide a data storage approach that makes use of structured query language or SQL [23]. The introduction of these databases dates back to the 1970s when data schemas were not as complicated as they are today. Moreover, storage was expensive and data archival incurred high costs. With the rise of social media platforms, the amount of data being stored about events, objects and people has risen exponentially. The use of data in this time and age is not just limited to data archival, but it also extends to frequent data retrieval and processing, in order to serve purposes like generation of real-time feeds [24] and customized advertisements [25], in addition to many others.

Owing to the complexity of information being processed and the need to treat multiple database requests to answer a single API request or render a webpage, the demands from modern database systems are ever-increasing. Some of the key drivers in this domain are the need for interactivity, increasing complexity and ever-evolving networks of users [26]. In order to serve these growing demands, sophisticated deployment strategies and improved computing infrastructure [188] are being put to use. With that said, single server deployments are expensive and highly complex, which has caused a drift towards the use of cloud hardware [27] for this purpose. Besides this, the use of agile methods has also reduced the development and deployment time [28], allowing quicker response to user needs.

Relational databases were not created to manage the agility and scalability requirements of modern-day systems. Moreover, they are also not equipped to work with the cloud and take optimum advantage of its cheaper storage and processing capabilities. These shortcomings can be addressed using two main technical approaches, which are discussed below:

- Manual Sharding

  In order to make use of the distributed paradigm, tables need to be segmented into smaller units, which must then be stored across different machines. This process of splitting is called manual sharding [29]. However, this functionality is not available in a traditional database and needs to be implemented by the developer. Moreover, the storage of data on each instance is performed in an anonymous mode.

  It is the responsibility of the application code to segment data, store it in a distributed manner, and perform query management and aggregate results to be presented to the user. Additional code shall be required for supporting data rebalancing, performing join operations, handling of resource failures and replication. It is crucial to mention that manual sharding may downgrade some of the benefits of relational databases like transactional integrity.

- Distributed Cache

  Caching [30] is a commonly used process, which is primarily employed for improving the read performance of a system. It is noteworthy that the use of a cache has no impact on the write performance and is capable of adding substantially to the complexity of the overall system. Therefore, if the requirements from the system are read-intensive, then the use of distributed cache must be considered. On the other hand, write-intensive or read/write intensive applications do not require a distributed cache [31].

NoSQL databases [13] are known to mitigate the challenges associated with traditional databases. In addition, they also unleash the true power of cloud by making use of commodity hardware, which reduces the cost, and simplifies deployment, making the life of a developer much easier as there is no need to maintain multiple layers of cache anymore. NoSQL is an umbrella term used to describe a plethora of technologies, all of which entail some common characteristics, which have been discussed later in this section. Some of the advantages of NoSQL solutions over traditional databases are as follows:

- Scalability

  NoSQL allows systems to scale out horizontally [13]. Moreover, this can be done quickly without affecting the overall performance of the system with the help of cloud technologies. Scaling traditional databases require manual sharding that involves high costs and complexity. On the other hand, NoSQL solutions offer automatic sharding, reducing complexity as well as cost of the system [13].

- Performance

  As mentioned previously, NoSQL systems can be scaled out as required. With the increase in the number of systems, the performance of a system is also correspondingly improved. The fact that these systems involve automatic sharding means that the overhead associated with the same is also eliminated, which further contributes to the improved performance of the system.

- High and Global Availability

  Relational databases depend on primary and secondary nodes to fulfill the availability requirements. This not only adds to the complexity of the system, but it also makes the system moderately available. On the contrary, NoSQL solutions make use of master-less architecture and data is distributed across multiple machines [13]. Therefore, even upon the failure of a node, the availability of the application remains unaffected for read as well as write operations.

  NoSQL solutions offer data replication across resources [13]. Consequently, user experience is consistent irrespective of the location of the user. Moreover, it also plays a significant role in reducing latency with the added advantage of shifting the developer's focus from database administration to business primacies.

- Flexible Data Modeling

  It is possible to implement fluid and flexible data models in NoSQL [13]. This allows developers to implement query options and data types that befit the application instead of those that suit the schema. In the process, the interaction between database and application is simplified, making this approach a better option for agile development.

### 3.1 Dynamic Schemas

Relational databases [33] have an inherent requirement to create schemas in advance. Data is added to the database only after this requirement is fulfilled. For instance, if a system needs to store employee data like name, department, age, gender and salary, then the table created for the same must have the corresponding schema. Such a requirement is unfit for agile development environments, as the fields of data might need to be changed over time. A new requirement may be added, as part of iteration, and subsequently, the schema may have to be altered. This is a time-consuming task if the database is large.

As a result, the database may have to be made unavailable for any use for a considerable amount of time to make required changes. Moreover, if the development process requires several iterations, the database may have to be shut rather frequently for significant amounts of time. Evidently, relational databases are inappropriate for storing data that are large, unstructured and unknown [33].

NoSQL satisfies this requirement since it has no predefined schemas. Moreover, data insertion does not require the developer to define a schema well in advance. As a result, changes to the data structure and data can be made in real-time without the need to shut the database for any other use [32]. There are several advantages of using this approach. Apart from the fact that it reduces administrator time, such an approach also reduces the time required for development and simplifies the process of code integration.

### 3.2 Auto-Sharding

Relational databases are structured in such a manner that they need to have a server that controls the rest of the systems to provide reliability and availability requirements of a database solution. Therefore, such a system can only support vertical scaling [32], which is not just expensive, but it leads to creation of small number of points of failure.

Besides this, it also places a limit on the amount of scaling that a system can support.

In view of the system requirements, a database solution must support horizontal scaling [13]. Therefore, it must be possible to add servers to the ensemble and get rid of the limitation that focuses on testing the capability of a single server. Cloud computing offers the best solution in this regard by providing on-demand services and unlimited scaling capacity [34]. So, the system no longer needs to rely on one server to fulfill its needs. Another important facet of using the Cloud is its inbuilt database administration facility. Moreover, the developer no longer needs to create complex platforms and can simply focus on writing the application code. Lastly, the use of Cloud-based, multiple servers cost significantly lesser than a high-capacity, single server.

In order to perform sharding of a database spanning across multiple servers, complex arrangements to make multiple servers act a single system, need to be put in place. On the other hand, NoSQL databases support auto-sharding. The database automatically distributes data across multiple systems without the need for the administrator to be aware of the server pool composition. Load balancing [35] for data and query are also automatically performed by the system. This allows the system to offer high availability. As and when the server goes down, it can be conveniently replaced and operations remain unaffected.

### 3.3 Automatic Replication

Replication is performed automatically for any NoSQL system [32]. Therefore, the system can recover from disasters rather easily, also allowing high degrees of availability. From the developer's point of view, he or she no longer needs to cater for these facets of development in the application code.

### 3.4 Integrated Caching

The integrated caching abilities [32] of NoSQL systems are rather well equipped and most of the frequently used data is kept in the system memory to ensure quick access. Therefore, there is no need to maintain multiple caching layers at the application level.

### 4.0 Classification Criteria for NoSQL Solutions

NoSQL is a technology that is developed to counter the issues presented by relational databases, which is implemented in multiple ways by different models. Common characteristics of NoSQL models include efficient storage, reduced operational costs, high availability, high concurrency, minimal management, high scalability and low latency [36]. NoSQL solutions have been classified using multiple criteria. The most commonly used classification criteria include supported data model (document-oriented, graph, key-value and wide-column) [41] and CAP characteristics (consistency, availability and partition tolerance) [63]. In view of the fact that the most commonly used classification criteria makes use of supported data model(s), big data models are discussed elaborately in this section. This section also includes CAP theorem and other relevant features and applications. These aspects of NoSQL solutions have been summarized in Table 2 for 80 identified NoSQL solutions.

### 4.1 Big Data Models

Yen [13] provided a detailed classification of NoSQL solutions dividing them into nine categories, which include Wide Columnar Store, Document Store, Object Database, Tuple Store, Data Structures Server, Key Value Store, Key-Value Cache, Ordered Key Value Store and Eventually Consistent Key Value Store. Another taxonomical study was performed by North [13], which gave a comprehensive classification and included cloud-based solutions as well for the analysis. Solutions have been classified under six categories namely Entity-Attribute-Value Data Stores, Amazon Platform Column Stores, Key-Value Data Stores and Distributed Hash Table Document Stores.

Cattell [39] and Leavitt [40] proposed a data model-based classification. Catel [39] divides NoSQL solutions into three categories namely key value stores, document stores and extensible record stores. On the other hand, Leavitt [40] proposed the use of three categories namely document-based, key value stores and column-oriented stores. Scofield [41] gave the most accepted categorization scheme by classifying databases into relational, graph, document, column and key value stores. This paper uses the above-mentioned basis for classification and covers document-oriented stores, graph data model, key value store and wide column store in this section.

### 4.1.1 Document-Oriented Data Model

Document-oriented data model uses documents for storage and retrieval of data. [42]. Other names of this NoSQL data model are document database [43] and document store [44]. It is primarily used for management of semi-structured data because of its flexibility and support for variable schema. As mentioned previously, document databases use documents for their working. These documents may be in PDF or Microsoft Word format. However, blocks of JSON and XML are the more commonly used document formats.

A relational database contains columns, which are described by their names and data types. On the contrary, in case of document databases, data type description and value for the concerned description are provided in a document [42]. The structure of different documents making up a database may be similar or different. Since, the document metadata does not store schema, there is no need to alter the metadata for adding distinct data elements to the database.

Documents are grouped together to form a structure called collection [42]. There may be multiple collections in a database. This structure is similar in functions to tables, which are present in relational databases [33]. Document-oriented databases provide a mechanism to execute queries on collections and retrieve documents that satisfy the attribute-specific requests. There are several advantages of using this approach, which include:

- Most of the growing data comes from IoT devices [302] and social media [303]. However, this data does not fit into standard application data models. Document-oriented databases offer flexible data modeling [42], in contrast to relational databases that force applications to fit data into existing models irrespective of their needs.

- The write performance of document-oriented databases is better than conventional systems [45]. In order to make a system available for writing, the database can compromise on data consistency as well. Therefore, even if a system fails and replication takes longer than expected, the write operation will be fast.

- The indexing features and query engines of databases available in this category are known to be fast and efficient [42]. Therefore, they offer better query performance.

Document databases work around documents. Therefore, a document is the basic atomic unit of storage in such databases. Any domain model that allows splitting and partitioning of its data across documents can use a document database. Some common use cases include CMS, blog-software and wiki-software [58]. However, when considering this data model, you may come across use cases where a relational model may be just as good an option to use as the non-relational database. The use cases are summarized in Table 1.

Table 1: Use Cases for Different Big Data Models

| Big Data Model | Preferred for Use Cases | Not Preferred for Use Cases |
|---|---|---|
| Document-Oriented [42] | 1. Content management systems<br>2. E-commerce platforms<br>3. Blogging platforms<br>4. Analytics platforms | 1. Applications requiring complex search queries.<br>2. Applications requiring complex transactions with multiple operations. |
| Key-Value [53] | 1. Storage of user preferences.<br>2. Maintenance of user profiles that don't have a specific schema.<br>3. Storage of session data for users.<br>4. Storage of shopping carts' data for multiple users. | 1. Specific data value needs to be queried.<br>2. Multiple unique keys need to be worked upon.<br>3. Frequent update of a part of the value.<br>4. Data values have established relationships with each other and the application requires exploitation of the same. |
| Graph [49] | 1. Network and IT operations<br>2. Graph based searches<br>3. Social networks<br>4. Fraud detection | Such a model is inappropriate for any application for which the data cannot be modeled as a graph. Transactional data, which is disconnected and in which relationship between data is not important, is an example. |
| Wide-Column [54] | 1. Blogging platforms<br>2. Content management systems<br>3. Counter-based systems | 1. Application requires complex querying.<br>2. Application has varying patterns of queries.<br>3. In scenarios where the database requirement is not |

| 4. Applications with write-intensive processing | established, the use of such a store must be avoided. |

### 4.1.2 Graph Data Model

This NoSQL data model is tailor made to support storage and processing of voluminous data, which may be semi-structured, structured or unstructured, in type. Therefore, data can be accessed and acquired from different sources. As a result, the graph data model [46] is popularly used in social media analytics [47] and different specialized fields of big data analytics [48]. It is noteworthy that relational databases were developed for storing structured information available in and generated by enterprises. Therefore, the schema of data to be stored is available beforehand. On the contrary, data generated by IoT (Internet of Things) and social media is unstructured. Moreover, it is generated in real time.

Graph databases [49] are a good option for storing unstructured data generated by such diverse sources at high velocity. There is no need to define a schema before storing data, which makes the database rather flexible. Besides this, graph databases are cost-effective and dynamic when it comes to integration of data coming from different sources [49]. Moreover, graph databases are better equipped to handle, store and process high-velocity data as compared to relational databases.

The aforementioned applications like social media analytics and IoT-based analytical solutions [187] require the base technology to integrate data coming from heterogeneous sources and establish links between the different datasets created. Application data of this kind can best be handled using a semantic graph database or RDF triplestore [50]. Semantic graph database is a type of graph database that focuses on relationships between different elements of the database and generate analytics on this basis. These graph databases are primarily used for real time analytics because of their ability to handle large datasets, without the need to define a schema in advance. The benefits of using semantic graph database can be summarized as follows:

1. Integration of inbound data from different sources is limited when the schema needs to be defined before adding data because the addition of a new source might require a change in schema, which is both time-consuming as well as complicated. In databases where there is no such need, data integration is limitless, simple and cost-effective [51].

2. Semantic graph databases offer an additional support to ontologies or semantically rich data schemas [51]. Therefore, organizations can create logical models in any way they desire.

3. Semantic graph databases use international standards for data representation on the web [51]. This results in easier integration and sharing of data. Uniform Resource Identifier (URI) [52] is one of the standards used for data representation in semantic graph databases. URI is a unique ID, which is used to distinguish between linked entities. The presence of such a clear approach for entity identification makes access and search easier, making the approach cost-effective. Moreover, it also makes data sharing easier as far as mapping data to Linked (Open) Data is concerned. In addition, challenges like vendor lock-in can be avoided.

Graph databases offer an effective way to manage and combine data. Enterprise data is typically linked and graph databases for their storage ensure easier management of content. Moreover, personalization can also be achieved in a simpler manner. In addition to this, the concept of connected world, which has particularly picked up pace after the rise of social media and IoT, can take advantage of the fact that graph databases allow integration of heterogeneous, interlinked data from different sources. The use cases corresponding to graph databases are summarized in Table 1.

### 4.1.3 Key-Value Data Model

Key-Value Store [53] implements schema-less policy by having no schema and making the data value opaque, which makes it the most flexible data model. The data value can store strings, numbers, images, binaries, counters, XML, JSON, HTML and videos, in addition to many others [53]. The stored values can be accessed with the help of a key. The flexibility of the database is manifested in the fact that the application controls the data value, completely. Key benefits of key-value stores are as follows:

1. The database does not force the application to structure its data in a specific form. Therefore, the

application is free to model its data in accordance with the requirements of the use case.

2. Objects can simply be accessed with the help of a key assigned to the object. When using this database, there is no need to perform operations like union, join and lock on objects [53], which make this data model, most efficient and high performing.

3. Most of the available key-value databases allow scale out as and when the demand for the same arises. Moreover, this can be done using commodity hardware without the need for any redesigning.

4. Providing high availability is much easier and uncomplicated with key-value stores. The distributed architecture and master-less configuration of some of the available databases of this type ensures higher resilience [53].

5. The design of these databases is such that it is simple to add and remove capacity. Moreover, these databases are better equipped to deal with network failures and hardware malfunctions [53], lowering the downtime considerably.

Key-value stores are commonly used data models, preferred for application areas surrounding data like user profile, emails, blog/article comments, session information, shopping cart data, product reviews, product details and Internet Protocol (IP) forwarding tables [59], in addition to many others. It is crucial to understand that a key-value store can be used to store complete webpages [60]. In this case, URL can be used as the key, and webpage content, as value. However, other data models may be better suited for this purpose if the application requires so. The use cases are mentioned in Table 1.

### 4.1.4 Wide-Column Data Model

Wide Column Stores [54] have columns and column families, as base entities. Facts or data are grouped together to form columns, which are further organized in the form of column families that are constructs similar to tables in relational databases. For example, data about an individual like name, account name and address are facts about the individual and can be grouped together to form a row in a relational database. On the contrary, same facts are organized in the form of columns in a wide-column store and each of the columns includes multiple groups. Therefore, a single wide-column can store data equivalent to the same stored by many rows in a relational database. Other names of such databases include column-oriented DBMS [55], columnar databases [56] and column families [57].

Key advantages [54] of using wide column store databases include:

1. Partitioning and data compression can be performed efficiently using wide column store databases.

2. Aggression queries like AVG, SUM and COUNT can be performed effectively and efficiently because of the column-oriented structure of this database.

3. This database type is highly scalable and well suited for massively parallel processing (MPP) systems.

4. Tables with huge amounts of data can be loaded and queried with relatively lesser response time.

Wide-column stores form the last category of big data models. These databases are deemed most appropriate for distributed systems [61]. In other words, if the data available is large and can be split across machines, then a wide-column store database can be extremely useful. Some of the primary advantages of using this database is reduced query time for some queries. However, this point must be clearly investigated before a decision in favor of such a solution is taken. For some queries, the time may be same or higher than that offered by conventional RDBMS solutions [62]. Typical use cases are recapitulated in Table 1.

## 4.2 CAP Theorem

While discussing the applicability of NoSQL solutions to real-world problems, it is important to mention CAP Theorem [63]. This theorem introduces the concept of Consistency (C), Availability (A) and Partition Tolerance (P) for distributed systems and states that all these three characteristics cannot be ensured by a solution simultaneously. In fact, a solution can provide at most two characteristics. Consistency is a characteristic that ensures that all the nodes of the distributed system must read the same value of data at all times. If a change in data value is made, then the change must be consistent for all nodes. However, if the change results in an error, then a rollback must be performed to ensure consistency.

Availability defines the operational requirement of the system that ensures that as and when a user makes a request to the system, it must respond to it despite its state. Partition tolerance refers to a system's ability to operate despite failure of a partition and message loss. It can also be described as the ability of a system to operate irrespective of network failure. Different database solutions and their CAP status have been described in the following sections. It has been stated that a distributed system can only possess two characteristics at a time. On the basis of this assertion, NoSQL systems can be CA (Consistent-Available), AP (Available-Partition Tolerant) or CP (Consistent-Partition Tolerant) [64].

## 4.3 Other Features

In addition to the above mentioned, there are many other features that can be considered for classification of NoSQL solutions. Some of these include ownership (free and proprietary), concurrency control, replication model, partitioning scheme, supported programming languages, compression support and indexing [291], among others. Out of these features, ownership has been considered in this paper. Other features can be considered in future work to enhance the proposed classification scheme (Section 6.0) and solution-application suitability model (Section 7.0).

## 5.0 Applications

Big data technologies are applicable to varied fields and domains. In view of this, 152 related resources were reviewed to determine the applications of different NoSQL solutions (mentioned in Table 2). Applications, in this context, can broadly be divided into the categories given below. The objective of this categorization was to determine popular application areas where NoSQL solutions have been put to use. This categorization has been used in Section 7.0 for further analysis.

1. Smart Cities
   Existing literature [296] suggests that smart cities is an application of Internet of Things (IoT) and includes applications that are focused towards improvement of quality of life of citizens involved. Considering the wide realm of this application area, a majority of the research studies reviewed belonged to this category. Applications include smart education [236, 222, 229, 154, 158, 170, 184, 278], intelligent waste management [255], smart agriculture [191], smart governance [282, 150], intelligent natural resources management [226, 70] and intelligent traffic control systems [257, 220, 148], in addition to many others.

2. Social Networks Analysis
   The rise of Internet has popularized social networking to such a high degree that this technological development has revolutionized the way people connect and communicate. With that said, this application area is one of the leading data generators with data being generated every second in huge volumes that includes not just textual data, but different forms of multimedia data as well. Research studies related to social network analysis include storage systems for social network data [213, 201], data management for social networks [214], processing [180], graph analysis [156], real-time processing [185] and analysis [241], and social intelligence applications[150, 277]. Papers related to microblogging and sentiment analysis have also been included in this category. Applications related to microblogging include data management [147] and text extraction and real-time sentiment analysis [216].

3. Geospatial Data Analysis
   Geo-data is another typically large dataset that needs to be analyzed to gain useful insights and make predictions that are crucial for mission critical applications and projects. The applications associated with this domain span across data representation [223], semantic data management [212], big geospatial raster data management [215] and geospatial/GIS applications [224, 264].

4. Life Sciences
   Life Sciences research remains one of the most impacted domain with the evolution of big data technologies as many applications like genomics and database maintenance are largely dependent on them in the modern scenario. In view of the fact that biologists and scientists working in this domain deal with mammoth-sized datasets that conventional systems fail to store and manage, big data technologies have reduced management effort and system response time considerably. Most of the papers identified for this application area were centered on database management, storage and analysis [198, 207, 204, 205, 152, 166, 206] with only one of the papers that elaborated upon comparative genomics application [227].

5. Healthcare

The rapid development in new-age technologies like Internet of Things (IoT), Cloud and Big Data has greatly impacted heathcare. The literature surveyed for the use of NoSQL technologies in healthcare included varied applications. Data is typically collected from IoT-based sensors and collected for medical applications [242]. Clinical databases [169] are being developed with specific focus on Electronic Medical Records (EMR) database [182]. Data analysis and monitoring applications include real-time analysis of Electrocardiogram (ECG) [176], medical imaging applications [182], breast imaging analysis [171], health information systems [208], mining of biomedical networks [284], biomedical applications [273], clinical big data analysis [261], specific applications for radiology [244]. Some problem-specific solutions like health danger prediction [258], prediction of health issues based on evaluation of toxicity [251, 252, 253, 254], patient safety application and estimation of calorific expenditure [163] were also identified.

6. Business Intelligence

These big data applications are focused towards improving decision making and operational efficiency of organizations. Applications associated with this domain include data management [218, 96, 280], problem-specific applications for areas like financial services [232] and service performance management [233], and general applications [69, 286, 150, 167, 181] for improving the operational efficiency of the system. It is important to mention that applications related to industries and company-specific applications like logs analysis for IT companies are not included in business intelligence. They are classified under 'Others' category.

7. Others

Many industry-specific applications like an application for proactive semiconductor equipment maintenance [199] and analytical solutions for construction [153] and insurance industry [209], have been classified under this category. Moreover, Internet of Things (IoT) applications that do not fall under the above mentioned categories have also been classified under this category. Lastly, task-specific systems like document management systems [245, 262, 263, 265, 269], indexing engine [202] and biometric system [250], in addition to many such others have also been put in this category.

## 6.0 Analysis of NoSQL Solutions

This paper identifies 80 NoSQL solution and presents a qualitative assessment of available literature for the same to determine the corresponding data model, CAP characteristics, ownership and existing applications, details of which have been described in the previous section. Table 2 provides a summary of this qualitative assessment. If a NoSQL solution supports the data model, CAP characteristics or ownership model, then the value corresponding to that cell was set to 1. On the other hand, in the absence of support, the value was reset to 0. Therefore, the dataset contains data of NoSQL solutions corresponding to 9 features, which are described below. It is noteworthy that creation of this dataset is purely done on the basis of literature review.

Quantitative analysis of this dataset was done using many techniques. This analysis is divided into two parts namely, bivariate analysis and cluster analysis, which have been discussed in the following sections. Fig. 2 shows a histogram of the frequency distribution of data for different features. Corresponding to this analysis, a base dataset was created with the following columns:

1. Document-Oriented – Indicates support for document-oriented data model
2. Graph - Indicates support for graph data model
3. Key-Value - Indicates support for key-value data model
4. Wide-Column - Indicates support for wide-column data model

5.  Consistent – Symbolic of consistency from CAP characteristics
6.  Available - Symbolic of availability from CAP characteristics
7.  Partition-Tolerant - Symbolic of partition tolerance from CAP characteristics
8.  Free – Represents if the solution is available free of cost or is open source
9.  Proprietary - Represents if the solution is available at a price or subscription

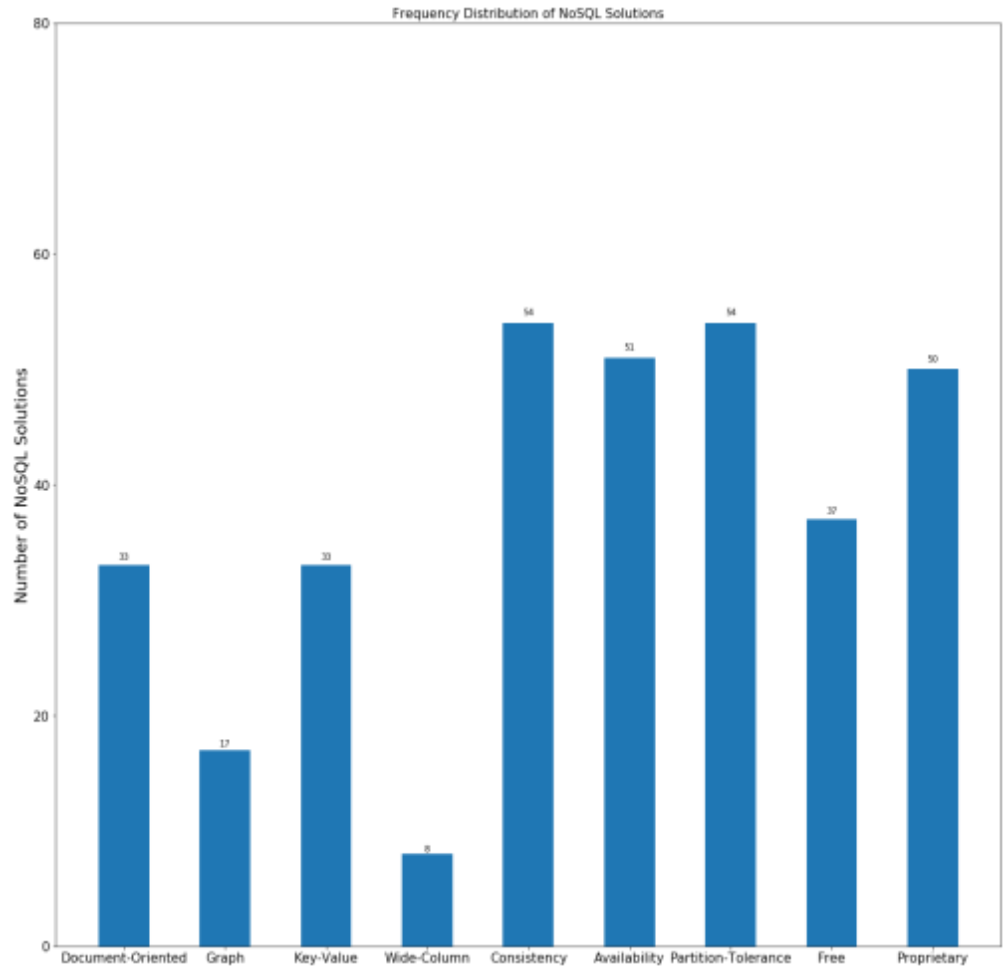

Fig. 2. Frequency Distribution Histogram for NoSQL Solutions Dataset

## 6.1 Bivariate Analysis

Bivariate analysis [295] between pairs of variables or features used in the dataset was performed to examine relationship between different features. In view of the fact that available data is categorical in nature, the techniques chosen were spearman's rank correlation [292] and the chi-square test [293]. Pearson correlation between the rank values of two variables is computed to determine the spearman's rank correlation. This measure is used for quantification of statistical dependence between two variables. The value of spearman's rank correlation coefficient lies in the range {-1, 1}. While the negative sign represents reciprocal association, a positive value is indicative of direct association. Values greater than 0.4, whether positive or negative, indicate moderate to strong association [292]. The heatmap for coefficient values between pairs of features is shown in Fig. 3. Computations performed for the given dataset show four moderate (negative) correlations. The feature-pair and coefficient value, greater than 0.4 (positive or negative), are provided in Fig. 3.

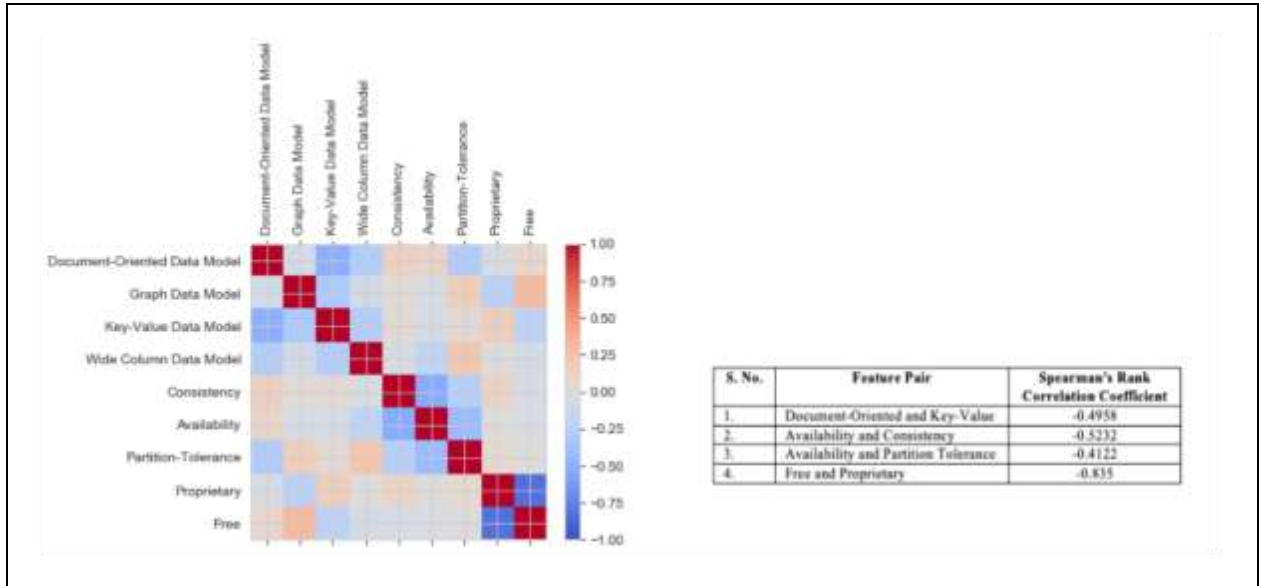

Fig. 3. Heatmap of Spearman's Rank Correlation Coefficient

In order to explore the existence of statistically significant relationships between different features, chi-square test was performed. The calculated p-values are compared with the threshold value of 0.05 [293]. If $p < 0.05$, null hypothesis is rejected and it is concluded that a relationship exists between the two variables. In other words, the value of one variable can help in predicting the value of the other variable and they can be referred to as 'dependent'. On the other hand, if the p-value is significantly high, the null hypothesis is confirmed and it is inferred that no relationship exists between the variables. The heatmap for p-values corresponding to pairs of features is shown in Fig. 4. Computations performed for the given dataset establish relationship between 9 pairs of features. The feature-pair and corresponding p-value for these 9 pairs is provided in Fig. 4.

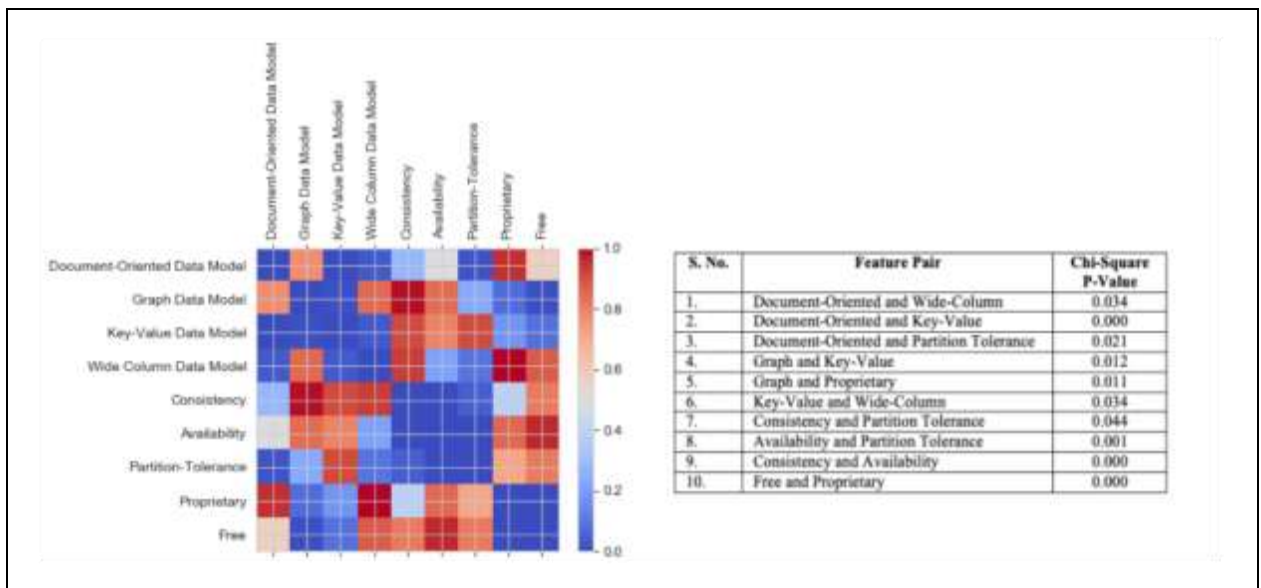

Fig. 4. Heatmap of Chi-Square P-Value

Table 2: Comparison of NoSQL Solutions

| S. No. | NoSQL Solution | Supported Data Models | CAP Characteristics | Other Features | Applications |
|---|---|---|---|---|---|
| 1 | AllegroGraph [65] | Document – Oriented and Graph | Consistent and Partition-Tolerant | 1. Proprietary<br>2. Supports RDF, JSON and JSON-LD<br>3. Provides Multi-Master Replication<br>4. Supports ACID Transactions, two-phase commit and full-text search | 1. Biological Database Creation [198]<br>2. Semantic Indexing Engine [202]<br>3. Context-driven analysis in cultural heritage environments [203]<br>4. Big Data Analytics for Insurance Industry [209]<br>5. Geospatial Semantic Data Management [212]<br>6. Data management and query handling for the social networks, Friendfeed [214] |
| 2 | Accumulo [66] | Wide-Column | Consistent and Partition-Tolerant | 1. Free<br>2. Scalable and distributed<br>3. It is built over and above Hadoop [8], Thrift [74] and Zookeeper [107]<br>4. Provides cell-level security and mechanisms for server-side programming | 1. Backend Storage for life sciences databases [207] |
| 3 | Aerospike [67] | Key-Value | Consistent and Partition-Tolerant | 1. Free<br>2. Highly scalable<br>3. Flash-optimized, in-memory<br>4. Reliable and consistent<br>5. Used for applications like dynamic web portals, user profiling and fraud detection | 1. Data management and query handling for the social networks, LinkedIn and Foursquare for deployment of Voldemort and Riak respectively [214] |
| 4 | Amazon Neptune [68] | Graph | Consistent and Partition-Tolerant | 1. Proprietary<br>2. Fully managed database<br>3. Provided as a web service<br>4. Supports RDF and property graph models<br>5. Supports SPARQL [196] and TinkerPop Gremlin [197] query languages | 1. Metadata repository for Smart Energy Services or Intelligent Energy Management [217]<br>2. Data Management Systems for BI Applications [218] |
| 5 | AnzoGraph [69] | Graph | Consistent and Partition-Tolerant | 1. Proprietary<br>2. Massively parallel<br>3. Graph Online Analytics Processing (GOLAP) database<br>4. Supports SPARQL [196] and Cypher [145]<br>5. Originally designed to analyze semantic triple data interactively | 1. Business Intelligence Applications [69] |
| 6 | ArangoDB [65] | Document – Oriented, Key-Value and Graph | Consistent and Available | 1. Free<br>2. Supports multiple database models with a single core<br>3. Possesses unified query language, called ArangoDB Query Language (AQL), which enables a single query for accessing multiple data stores. | 1. Android application for controlling traffic [220]<br>2. Progressive web application that maintains a database of coins and historical figures [139]<br>3. Predictive analytics involving collaboration platform between academia and industry [222]<br>4. Representation and enrichment of Geo-data [223] |

| 7 | Azure Tables [71] | Wide Column | Available and Partition-Tolerant | 1. Proprietary<br>2. Provisioned as a service where storage of data is allowed into collections that can be partitioned. Data is accessed by means of primary and partition keys. | 1. GIS Application [224]<br>2. Processing of Seismograms [225]<br>3. Ground water simulation and related applications [226]<br>4. Comparative Genomics applications [227] |
|---|---|---|---|---|---|
| 8 | BaseX [72] | Document-Oriented | Consistent and Available | 1. Free<br>2. Provides support for JSON, XML and binary formats<br>3. Implements master-slave architecture<br>4. Provides support for concurrent structural and full-text update/search | 1. Annotation and retrieval in digital humanities research [228] |
| 9 | BerkeleyDB [73] | Key-Value | Consistent and Partition-Tolerant | 1. Free (with commercial versions also available)<br>2. High performing and scalable<br>3. Supports complex data management<br>4. Most appropriate for applications requiring embeddable database. | 1. E-learning system [229]<br>2. Collaborative web search [230] |
| 10 | BigTable [75] | Wide-Column | Consistent and Partition-Tolerant | 1. Proprietary<br>2. High performance<br>3. Provides data compression | 1. Management of sensor network data [231]<br>2. Financial services application like audit trail [232] |
| 11 | Cache [76] | Document – Oriented | Available | 1. Proprietary<br>2. Data is stored in multi-dimensional arrays. Therefore, structured data that is hierarchical in nature can be stored.<br>3. Commonly used for business and health-related applications | 1. Service performance management [233]<br>2. EMR database [182] |
| 12 | Cassandra [77] | Wide-Column | Available and Partition-Tolerant | 1. Free<br>2. Distributed<br>3. Highly available<br>4. Master-less replication with robust support for clusters across multiple datacenters | 1. Analytics for proactive semiconductor equipment maintenance [199]<br>2. Context-driven analysis in cultural heritage environments [203]<br>3. Storage and management of Life Sciences Databases using CumulusRDF, which performs linked data management on nested key-value stores [204]<br>4. Management of data and query handling for the social networks, Facebook, Friendfeed, Foursquare and Twitter [214]<br>5. IoT applications [79] |
| 13 | CDB or Constant Database [78] | Key-Value | Consistent and Available | 1. Free library<br>2. On-disk associative array that maps keys to values, allowing a key to have multiple values<br>3. It can be used as a shared library. | During literature review, no applications were found for this NoSQL solution. |
| 14 | Cloudant [68] | Document-Oriented | Available and Partition-Tolerant | 1. Proprietary<br>2. Distributed database service<br>3. Uses BigCouch [189] and JSON model, at backend | 1. Environmental sensing applications [234]<br>2. Application that finds and ranks researchers [235]<br>3. Smart campus application [236] |
| 15 | Clusterpoint Database [80] | Document-Oriented and | Consistent and Partition- | 1. Proprietary<br>2. Distributed JSON/XML database platform | 1. Medical Database Analysis [237]<br>2. Enterprise content management [238] |
| | | Key-Value | Tolerant | 3. Transactions are compliant with ACID properties<br>4. Highly available<br>5. Provides sharding and replication<br>6. Uses SQL or JS as query language | |
| 16 | Coherence [81] | Key-Value | Consistent and Available | 1. Proprietary<br>2. In-memory data grid and distributed cache<br>3. Appropriate for systems requiring high scalability and availability keeping latency at lower levels | 1. IoT-based analytical applications [239] |
| 17 | CouchBase Server [82] | Document-Oriented and Key-Value | Consistent and Partition-Tolerant | 1. Free<br>2. Distributed database<br>3. Uses SQL as querying language<br>4. Uses JSON model | 1. Traffic forecasting application that works in real time for environments, which implement the fog computing concept [240]<br>2. Social network analytics [241]<br>3. Collection of sensor data for medical applications [242]<br>4. Decentralized social networking application [243] |
| 18 | CouchDB [83] | Document-Oriented | Available and Partition-Tolerant | 1. Free<br>2. Supports JSON over HTTP/REST<br>3. Provides limited support for ACID transactions<br>4. Supports multi-version concurrency control | 1. Data management and query handling for the social networks, Foursquare [214]<br>2. Analytical solutions for radiology [244]<br>3. Document management system for software projects [245]<br>4. Querying system for graphical music documents [246]<br>5. Web-based application for monitoring and visualizing energy consumption of a house for improving energy efficiency [247] |
| 19 | CrateIO [84] | Document-Oriented | Available and Partition-Tolerant | 1. Free<br>2. It is based on the Elasticsearch/Lucene ecosystem<br>3. Supports objects that are binary or are also called BLOBs.<br>4. Makes use of SQL syntax for distributed querying of the system in real time | 1. Industrial IoT applications [248] |
| 20 | CosmosDB [85] | Document-Oriented | Available and Partition-Tolerant | 1. Proprietary<br>2. Provisioned as Platform-as-a-Service<br>3. Based on DocumentDB [87] | 1. Applications like assessment of risk for nuclear power plants [249]<br>2. Iris-based biometric system [250]<br>3. Screening of chemicals for liver toxicity [251], dose toxicity [252, 254] and non-cancer threshold of toxicological concern [253]<br>4. Smart city applications like automated waste management [255], traffic monitoring system [257] and automated ticketing system for public transport [256]<br>5. Prediction of dangers related to human health using IoT-based framework [258] |
| 21 | DataStax Enterprise Graph [86] | Graph | Available and Partition- | 1. Proprietary<br>2. Scalable, distributed database | 1. Data Management Systems for BI Applications [218] |

| | | | Tolerant | 3. Allows real-time querying<br>4. Supports Tinkerpop [197]<br>5. It is known to integrate well with Cassandra [77] | 2. Analytics for center of solar system [259]<br>3. Covert network analysis applications [260]<br>4. Clinical big data applications [261] |
|---|---|---|---|---|---|
| 22 | DocumentDB [87] | Document-Oriented | Consistent and Partition-Tolerant | 1. Proprietary<br>2. Provisioned as a database service<br>3. Fully managed version of MongoDB | 1. Document management system for patent application related documents [262]<br>2. Paper-digital document management system [263]<br>3. Geospatial applications [264]<br>4. User-based document management system [265] |
| 23 | Dynamo [88] | Key-Value | Available and Partition-Tolerant | 1. Proprietary<br>2. Distributed datastore<br>3. Highly available<br>4. Supports incremental scalability, symmetry among nodes, decentralization and it exploits the heterogeneity of the infrastructure it works on. | 1. Data management and query handling for the social networks, MySpace [214] |
| 24 | ElasticSearch [89] | Document-Oriented | Consistent and Available | 1. Free<br>2. Supports JSON<br>3. Basically a search engine | 1. External Indexing Engine for Smart City Applications [210]<br>2. Querying system for graphical music documents [246] |
| 25 | etcd [68] | Key-Value | Available and Partition-Tolerant | 1. Free<br>2. Supports binary data<br>3. Allows versioning, validation, collections, triggers, clustering, Lucene full text search, ACLS and XQuery Update<br>4. Uses XML over REST/HTTP | 1. Telecommunication applications [266]<br>2. Time critical applications [267]<br>3. Traffic forecasting in real-time for computing infrastructures [240] |
| 26 | eXist [91] | Document-Oriented | Consistent and Available | 1. Free<br>2. Supports text, JSON, HTML and XML formats, in addition to binary formats<br>3. XQuery is the provided querying language while XSLT is the corresponding programming language | 1. Document management system for linguistic applications [269]<br>2. Search applications on Bilingual lingual digital libraries [318] |
| 27 | FoundationDB [92] | Key-Value | Consistent, Available and Partition-Tolerant | 1. Free<br>2. Complies with ACID properties<br>3. Scalable<br>4. Allows replications<br>5. Bindings for Python, C, PHP and Java, in addition to many other programming languages is available | 1. IoT applications [270] |
| 28 | GridGain Systems [93] | Key-Value | Consistent and Partition-Tolerant | 1. Proprietary<br>2. Services and software solutions are provided for systems dealing with big data<br>3. Supports in-memory computing<br>4. Provides improved throughput and reduced latency | 1. Text documents' classification [271] and clustering [272] |
| 29 | GT.M [54] | Key-Value | Available | 1. Free<br>2. Developed for transaction processing<br>3. Supports ACID transactions | 1. Biomedical applications like electronic health records [273] |

| | | | | 4. Supports replication and database encryion | |
|---|---|---|---|---|---|
| 30 | Hazelcast [95] | Key-Value | Available and Partition-Tolerant | 1. Free<br>2. MapStore can be defined by the user<br>3. MapStore can be persistent<br>4. High consistency and supports sharing in the form of consistent hashing | 1. Bug prediction system for GitHub projects database [274]<br>2. Prediction of performance of IoT applications [275] |
| 31 | HBase [4] | Wide-Column | Consistent and Partition-Tolerant | 1. Free<br>2. Distributed database<br>3. It runs on top of Hadoop and provides capabilities similar to that of BigTable.<br>4. It is fault-tolerant for scenarios where a large amount of sparse data is being dealt with. | 1. Analytics for proactive semiconductor equipment maintenance [199]<br>2. SPARQL query engine, Jena HBase, for life sciences databases [205]<br>3. Data management and query handling for the social networks, Facebook, Twitter, Friendfeed and LinkedIn [214]<br>4. Real-Time Sentiment Analysis and Text Extraction for microblogging applications [216] |
| 32 | Hibari [96] | Key-Value | Available and Partition-Tolerant | 1. Free<br>2. Distributed big data store<br>3. Highly available<br>4. Strongly consistent | 1. Classification and clustering of visual online information [276]<br>2. Social big data applications [277]<br>3. Management of data for digital economy [96] and business intelligence [280]<br>4. Data collection and classification from wireless sensor networks [279] |
| 33 | HyperGraphDB [97] | Graph | Available and Partition-Tolerant | 1. Free<br>2. Schemas are dynamic and flexible<br>3. Knowledge representation and data modeling are efficient<br>4. Non-blocking concurrency<br>5. Appropriate for semantic web and arbitrary graph use cases | 1. Document-centric information systems' analysis [281]<br>2. E-government applications like citizen relationship management [282]<br>3. Intelligent manufacturing applications [283]<br>4. Mining of biomedical networks [284]<br>5. Applications in Governance [285] |
| 34 | HyperTable [98] | Wide-Column | Consistent and Partition-Tolerant | 1. Proprietary<br>2. It is based on BigTable<br>3. Massively scalable | 1. Business intelligence applications [286]<br>2. Applications in digital forensics [287]<br>3. Smart applications like online vehicle tracking system [288] |
| 35 | IBM Informix [99] | Document-Oriented | Consistent and Available | 1. Proprietary<br>2. RDBMS that supports JSON<br>3. Complies with ACID rules<br>4. Supports sharding and replication | 1. Enterprise applications for business intelligence [289] |
| 36 | IBM Informix C-ISAM [100] | Key-Value | Consistent and Available | 1. This API complies with Open Standards [304]<br>2. Allows management of data files, which have been organized using B+ indexing<br>3. It is the file storage used by Informix [99]. | 1. Development of effective handheld solutions [290] |
| 37 | Ignite [101] | Key-Value | Consistent, Available and | 1. Free<br>2. Distributed, in-memory computing platform | 1. Data management of microblogs [147]<br>2. Data analysis of live traffic for intelligent city |

| | | | | 3. Provides caching and processing platform<br>4. Provides support for ACID transactions and MapReduce jobs<br>5. Allows partitioning, clustering and replication<br>6. Highly consistent | traffic management systems [148] |
|---|---|---|---|---|---|
| 38 | InfiniteGraph [102] | Graph | Consistent and Partition-Tolerant | 1. Proprietary<br>2. Cloud-enabled<br>3. Distributed<br>4. It is scalable and cross-platform<br>5. It is capable of handling high throughput | 1. Smart Grid applications [149]<br>2. Applications for business, social and government intelligence [150] |
| 39 | InfinityDB [103] | Key-Value | Consistent, Available and Partition-Tolerant | 1. Proprietary<br>2. Completely developed in Java and includes DBMS and database engine<br>3. Based on B-tree architecture<br>3. Provides high performance<br>4. Reduces risks associated with failures | 1. On-device database for mobiles [151] |
| 40 | Jackrabbit [104] | Document-Oriented | Consistent and Available | 1. Free<br>2. Implementation of Java Content Repository | 1. Data management of biological investigations of systems biology [152]<br>2. Database of work-related accidents in construction industry [153] |
| 41 | JanusGraph [105] | Graph | Available and Partition-Tolerant | 1. Free<br>2. Distributed<br>3. Scalable and integrates well with backend databases like HBase [4], Cassandra [77], BigTable [75] and BerkleyDB [73]<br>4. Integrates well with platforms like Giraph [144], Spark [8] and Hadoop [12]<br>5. Provides support for full text search by external integration with Solr [137] and Elasticsearch [89] | 1. Exploration of scholarly networks [154, 158]<br>2. Smart City applications [155]<br>3. Analysis of social graph data [156]<br>4. Applications like network security analytics [157] |
| 42 | KAI [106] | Key-Value | Available and Partition-Tolerant | 1. Free<br>2. Scalable<br>3. Highly fault-tolerant<br>4. Provides low latency<br>5. Used for social networks and web repositories | 1. Analytical applications for IoT [239] |
| 43 | LevelDB [108] | Key-Value | Consistent and Partition-Tolerant | 1. Free<br>2. Maintains byte arrays for storing key and value pairs.<br>3. Data compression is supported by means of Snappy<br>4. Supports forward/backward iteration and batch writing<br>5. Used as a library | 1. Analytical applications for IoT [160]<br>2. Web-based system to explore tourist network in New Delhi [161]<br>3. Mission critical applications like call for fire [162]<br>4. Secondary storage for blockchain [94] |
| 44 | Lightening Memor1-Mapped Database (LMDB) [109] | Key-Value | Consistent and Available | 1. Free<br>2. Embedded database<br>3. High performance<br>4. Provides API bindings for many programming languages.<br>5. Employs multi-version concurrency control is offers high | 1. Application that provides an estimate of an individual's calorific expenditure [163]<br>2. Database of violent audio-video content [164]<br>3. Aerial sensing applications that process image data in real time [165] |
| | | | | levels of reliability | 4. Searching, mapping and visualizing bioinformatics identifiers and keywords [166] |
| 45 | Lotus Domino [110] | Document-Oriented | Consistent and Available | 1. Proprietary 2. It is a multi-value database [54] | 1. Business intelligence applications [167] 2. Application to ensure safety of patients by improving supervision [168] 3. Development of Clinical database for varied purposes [169] 4. Decision support system for management of graduated student employment [170] |
| 46 | Marklogic [111] | Document-Oriented and Graph | Consistent, Available and Partition-Tolerant | 1. Free 2. Supports XML, JSON and RDF triples 3. Distributed 4. Provides high availability, full-text search, ACID compliance and security | 1. Breast Imaging applications [171] 2. Search applications on Bilingual lingual digital libraries [172] |
| 47 | Memcached [112] | Key-Value | Consistent and Partition-Tolerant | 1. Free 2. Memory caching system that is general purpose and distributed 3. Scalable architecture 4. Supports sharding | 1. Filesystem for eScience applications [173] 2. Used on Wikipedia backend to reduce load on database [174] 3. Used in call centers to alleviate load on database [175] |
| 48 | MemcacheDB [113] | Key-Value | Consistent and Partition-Tolerant | 1. Free 2. A version of memcached that has persistence 3. It is a memory caching system that is distributed and general purpose. 4. Development has halted on this solution. | 1. Applications related to IoT [160] |
| 49 | Microsoft SQL Server [114] | Graph | Consistent and Available | 1. Proprietary 2. Typically used for modeling many-to-many relationships between data 3. Integration of relationships are done into Transact-SQL and the foundation DBMS is SQL Server | 1. Management of scientific database and related applications [177] |
| 50 | MongoDB [6] | Document-Oriented | Consistent and Partition-Tolerant | 1. Free 2. Supports BSON or binary JSON [305] 3. Allows replication and sharding | 1. Biological Database Creation [198] 2. Analytics for proactive semiconductor equipment maintenance [199] 3. Storage of Real-time Data from Sensors [200] 4. Context-driven analysis in cultural heritage environments [203] 5. Data Storage for Smart City Applications [210] 6. Data management and query handling for the social networks, LinkedIn, Flickr, Foursquare and MySpace [214] 7. Handling of Big Geospatial Raster Data [215] 8. Secondary storage for blockchain [94] |
| 51 | MUMP Database [115] | Document-Oriented | Available | 1. Proprietary 2. MUMPS is a programming language with inbuilt database | 1. Automated management system for dairy farms [178] |

| | | | | 3. Used for applications related to health sector | 2. Healthcare applications [179, 180] |
|---|---|---|---|---|---|
| 52 | Neo4j [116] | Graph | Available and Partition-Tolerant | 1. Free<br>2. Can be used for ACID transactions<br>3. Supports clustering and high availability<br>4. Provides complete administrative support<br>5. Provides inbuilt REST API for interface with other programming languages | 1. Analytics for proactive semiconductor equipment maintenance [199]<br>2. Storage model for healthcare information systems [208]<br>3. Storage of social network data [213]<br>4. Data management and query handling for the social networks, Facebook, Twitter, Flickr and MySpace [214]<br>5. Metadata repository for Smart Energy Services or Intelligent Energy Management [217]<br>6. Representation and enrichment of Geodata [223]<br>7. Bug prediction system for GitHub projects database [274]<br>8. Application in supply chain management [159] |
| 53 | NoSQLz [117] | Key-Value | Consistent | 1. Proprietary<br>2. Complies with ACID properties<br>3. Allows CRUD (Create, Read, Update, Delete) operations<br>4. Easy to implement | - |
| 54 | ObjectDatabase++ [118] | Document-Oriented | Consistent and Available | 1. Proprietary<br>2. Binary<br>3. Structure of native C++ class | - |
| 55 | OpenLink Virtuoso [119] | Document-Oriented, Graph and Key-Value | Consistent and Partition-Tolerant | 1. Proprietary<br>2. Hybrid of database engine and middleware<br>3. High performance and secure<br>4. Supports SQL [23] and SPARQL [196] for performing operations on SQL tables and RDF<br>5. JSON, XML and CSV document types are supported | 1. Big Data Analytics for Insurance Industry [209]<br>2. Geospatial Semantic Data Management [212]<br>3. Metadata repository for Smart Energy Services or Intelligent Energy Management [217] |
| 56 | Oracle NoSQL Database [120] | Key-Value | Consistent and Available | 1. Proprietary<br>2. Supports horizontal scalability and transparent load balancing<br>3. Supports replication and sharding<br>4. It is highly available and fault-tolerant | 1. Business intelligence applications [181]<br>2. Analysis of medical imaging data [182] |
| 57 | Oracle Spatial and Graph [121] | Graph | Available and Partition-Tolerant | 1. Proprietary<br>2. Capable for handling RDF and property graphs | 1. Big Data Analytics for Insurance Industry [209]<br>2. Data Management Systems for BI Applications [218] |
| 58 | OrientDB [122] | Document-Oriented and Graph | Consistent, Available and Partition-Tolerant | 1. Free<br>2. Supports JSON over HTTP<br>3. Supports SQL-type language use<br>4. Can be used for ACID transactions<br>5. Supports sharding, multi-master replication, security features and schema-less modes | 1. Community detection and related applications [183]<br>2. Smart education application [184]<br>3. Bug prediction system for GitHub projects database [274]<br>4. Real time social networking applications [185] |

| 59 | PostgreSQL [123] | Document-Oriented | Consistent and Available | 1. Free<br>2. Supports JSONB, JSON function and JSON store<br>3. Supports HStore 2 and HStore [192] | 1. Storing configuration parameters and catalogs for Context-driven analysis in cultural heritage environments [203]<br>2. IoT Application [275] |
|---|---|---|---|---|---|
| 60 | Project Voldemort [124] | Key-Value | Available and Partition-Tolerant | 1. Free<br>2. Supports horizontal scalability<br>3. Availability is high for read/write operations<br>4. Fault recovery is transparent<br>5. Supports automatic partitioning and replication<br>6. Considered appropriate for applications with read-intensive operations | 1. Data management and query handling for the social networks, LinkedIn [214]<br>2. Applications related to IoT [160] |
| 61 | Qizx [103] | Document-Oriented | Consistent and Available | 1. Proprietary<br>2. Distributed XML database<br>3. Supports text, JSON and binaries<br>4. Provides integrated full text search | 1. Applications of the linguistic domain [186] |
| 62 | RavenDB [126] | Document-Oriented | Consistent, Available and Partition-Tolerant | 1. Free<br>2. Fully transactional and high performance<br>3. Highly available<br>4. Multi-platform and easy to use<br>5. Multi-model architecture that allows it to work well with SQL systems | 1. Storage system for social networks [201] |
| 63 | Redis [127] | Key-Value | Consistent and Partition-Tolerant | 1. Free<br>2. Read/write operations and access to data are efficient<br>3. Fault-tolerant<br>4. Supports automatic partitioning<br>5. Appropriate for applications involving structured strings | 1. Context-driven analysis in cultural heritage environments [203] |
| 64 | RethinkDB [128] | Document-Oriented | Consistent and Partition-Tolerant | 1. Free<br>2. Distributed database<br>3. Supports JSON<br>4. Provides sharding and replication | 1. Data analytical applications for IoT-based green house system [219]<br>2. Application that implements an agent-based Platform for Autonomous Sailing [221]<br>3. Real time data management of aquarium data [268]<br>4. Database for learning environment [278]<br>5. IoT-based Real time ECG monitoring for healthcare applications [176]<br>6. Agriculture IoT application [191] |
| 65 | Riak [45] | Key-Value | Available and Partition-Tolerant | 1. Free<br>2. Highly available and fault-tolerant<br>3. Highly scalable and easy to operate<br>4. Cloud storage and enterprise versions of Riak are also available<br>5. It supports automatic data distribution and replication for resilience and improved performance. | 1. Applications based on transactional services like Automatic Vehicle Location System [102]<br>2. Data management and query handling for the social networks, Foursquare [214] |
| 66 | RocketU2 [130] | Document-Oriented | Consistent | 1. Proprietary<br>2. Provides dynamic support | - |

| | | | | 3. Scalable<br>4. Reliable and efficient<br>5. Appropriate for business information management | |
|---|---|---|---|---|---|
| 67 | RocksDB [131] | Key-Value | Consistent and Partition-Tolerant | 1. Free<br>2. Embedded database that assures high performance<br>3. Supports all the features of LevelDB [108]. In addition, it also supports geospatial indexing, universal compaction, column families and transactions. | 1. Facebook's real time data processing [190]<br>2. Matric computation for big data applications [129]<br>3. Tagging system for blockchain analysis [125]<br>4. Blockchain-based Library circulation system [90] |
| 68 | SAP HANA [132] | Document-Oriented and Graph | Consistent and Available | 1. Proprietary<br>2. Supports JSON only<br>3. Can be used for ACID transactions | 1. Web-scale data management for business applications [211] |
| 69 | Scalaris [133] | Key-Value | Consistent and Partition-Tolerant | 1. Free<br>2. Highly available and fault-tolerant<br>3. Offers high scalability<br>4. Consistent<br>5. Self-managing<br>6. Minimal maintenance overhead<br>7. Considered appropriate for applications that are read/write intensive | 1. Applications related to IoT [160] |
| 70 | ScyllaDB [134] | Wide-Column | Available and Partition-Tolerant | 1. Free<br>2. Distributed<br>3. Designed for integration with Cassandra for reducing latency and improving throughput<br>4. It supports Thrift and CQL, protocols also supported by Cassandra | 1. IoT applications [79]<br>2. Biogas data analytics [70]<br>3. Logs analysis for IT establishments [37] |
| 71 | Sedna [135] | Document-Oriented | Consistent and Available | 1. Free<br>2. XML database | 1. IoT applications [239] |
| 72 | SimpleDB [136] | Document-Oriented | Available and Partition-Tolerant | 1. Proprietary<br>2. Distributed database<br>3. Used as web service with Amazon EC2 [193] and S3 [194]<br>4. Provides availability and partition tolerance | 1. Large-scale RDF Store development and management for life sciences databases [206] |
| 73 | Solr [137] | Document-Oriented | Available and Partition-Tolerant | 1. Free<br>2. Search engine written in Java<br>3. Supports real-time indexing, full text search, database integration, dynamic clustering and rich document handling<br>4. Provides index replication and distributed search<br>5. Scalable and fault tolerant | 1. Creation of digital libraries with the help of web crawl [38] |
| 74 | Sparksee [138] | Graph | Consistent and Partition-Tolerant | 1. Proprietary<br>2. Scalable<br>3. High performance<br>4. First graph database for mobile devices<br>5. Bindings available for C++, C#, Objective C, Python and Java | 1. Social networking applications [21]<br>2. Smart City applications [22]<br>3. IoT applications [239] |

| 75 | Sqrrl [139] | Graph | Consistent and Partition-Tolerant | 1. Proprietary<br>2. Distributed<br>3. Mass-scalable<br>4. Real-time database<br>5. Provides cell-level security | 1. Progressive web application that maintains a database of coins and historical figures [139] |
|----|-------------|-------|-----------------------------------|---------|---------|
| 76 | Tarantool [140] | Key-Value | Consistent and Available | 1. Free<br>2. Provides crash resistance with the help of maintenance of write ahead logs<br>3. Can be integrated with other applications and frameworks written in different programming languages. | 1. Corpus management system [15]<br>2. IoT applications [16] |
| 77 | TokuMX [141] | Document-Oriented | Consistent and Partition-Tolerant | 1. Free<br>2. Version of MongoDB [6]<br>3. Supports fractal tree indexing [195] | 1. Server-based monitoring of energy storage systems like Lithium-Ion battery packs for mobile as well as stationary applications [17] |
| 78 | TerraStore [143] | Document-Oriented | Consistent and Partition-Tolerant | 1. Free<br>2. In-memory storage<br>3. Dynamic cluster configuration<br>4. Persistent<br>5. Supports load balancing and automatic data redistribution<br>6. Used for structured big data | 1. IoT applications [239]<br>2. Storage system for social networks [331] |
| 79 | Tokyo Cabinet and Kyoto Cabinet [142] | Key-Value | Available and Partition-Tolerant | 1. Free<br>2. Provides two libraries for database management<br>3. Storage is done via hash tables and B+ trees<br>4. Provides limited support for transactions | 1. Applications related to IoT [160] |
| 80 | XAP [146] | Key-Value | Available and Partition-Tolerant | 1. Proprietary<br>2. Software platform for in-memory computing<br>3. Appropriate use cases for this solution include real-time analytics and transaction processing requiring low latency and high-performance levels. | 1. Business intelligence applications [181] |

## 6.2 Cluster Analysis

In the previous section, interdependency of different features was analyzed. Statistically, it can be seen that relationships exist between different features and none of the existing classification schemes are independent enough to classify NoSQL solutions. Therefore, a combination of features must be used to create discrete categories for classification. In order to create categories, k-modes clustering [294] technique is used considering the fact that the dataset contains purely categorical data. In contrast to k-means clustering that clusters data points on the basis of Euclidean distance between them, k-modes clustering technique forms clusters depending on matching category values for different data points in the cluster [297]. The implementation of k-modes clustering makes use of Cao's initialization scheme [298]. Moreover, as there is no established scheme for deciding the number of clusters, data distribution across clusters was used to decide this value.

Table 3: Data distribution for different cluster counts

| n = 3 | C0 | C1 | C2 | | | | | |
|-------|----|----|----|----|----|----|----|----|
| All | 37 | 28 | 15 | | | | | |
| DMCAP | 37 | 20 | 23 | | | | | |
| CAPFP | 49 | 18 | 13 | | | | | |
| DMFP | 52 | 25 | 3 | | | | | |
| **n=4** | **C0** | **C1** | **C2** | **C3** | | | | |
| All | 37 | 18 | 14 | 11 | | | | |
| DMCAP | 32 | 25 | 13 | 10 | | | | |
| CAPFP | 46 | 18 | 13 | 3 | | | | |
| DMFP | 39 | 22 | 3 | 16 | | | | |
| **n=5** | **C0** | **C1** | **C2** | **C3** | **C4** | | | |
| All | 26 | 12 | 20 | 13 | 9 | | | |
| DMCAP | 32 | 20 | 13 | 10 | 5 | | | |
| CAPFP | 45 | 18 | 13 | 3 | 1 | | | |
| DMFP | 38 | 22 | 3 | 16 | 1 | | | |
| **n=6** | **C0** | **C1** | **C2** | **C3** | **C4** | **C5** | | |
| **All** | **22** | **18** | **15** | **11** | **9** | **5** | | |
| DMCAP | 29 | 15 | 13 | 9 | 5 | 9 | | |
| CAPFP | 43 | 18 | 13 | 3 | 1 | 2 | | |
| DMFP | 30 | 22 | 3 | 16 | 1 | 8 | | |
| **n=7** | **C0** | **C1** | **C2** | **C3** | **C4** | **C5** | **C6** | |
| All | 24 | 14 | 11 | 11 | 9 | 5 | 6 | |
| DMCAP | 29 | 15 | 13 | 9 | 3 | 9 | 2 | |
| CAPFP | 41 | 18 | 13 | 3 | 1 | 2 | 2 | |
| DMFP | 28 | 22 | 3 | 14 | 1 | 8 | 4 | |
| **n=8** | **C0** | **C1** | **C2** | **C3** | **C4** | **C5** | **C6** | **C7** |
| All | 22 | 14 | 11 | 11 | 6 | 5 | 6 | 5 |
| DMCAP | 28 | 15 | 13 | 9 | 3 | 9 | 2 | 1 |

| | C0 | C1 | C2 | C3 | C4 | C5 | C6 | C7 | C8 |
|---|---|---|---|---|---|---|---|---|---|
| CAPFP | 26 | 18 | 13 | 3 | 1 | 2 | 2 | 15 | |
| DMFP | 26 | 18 | 13 | 3 | 1 | 2 | 2 | 15 | |
| **n=9** | **C0** | **C1** | **C2** | **C3** | **C4** | **C5** | **C6** | **C7** | **C8** |
| All | 22 | 14 | 11 | 10 | 6 | 5 | 6 | 5 | 1 |
| DMCAP | 27 | 15 | 13 | 9 | 3 | 9 | 2 | 1 | 1 |
| CAPFP | 15 | 18 | 12 | 3 | 1 | 2 | 2 | 15 | 12 |
| DMFP | 15 | 18 | 12 | 3 | 1 | 2 | 2 | 15 | 12 |

Table 3 provides insights into the distribution of data across clusters for different cluster counts. In the table, n is the number of clusters. 'All' represents execution results when all the 9 features were considered for clustering. 'DMCAP' represents execution results when 7 features (Document-oriented, Graph, Key-Value, Wide-Column, Consistency, Availability and Partition Tolerance) were considered for clustering. 'CAPFP' represents execution results when 5 features (Consistency, Availability, Partition Tolerance, Free and Proprietary) were considered for clustering. 'DMFP' represents execution results when 6 features (Document-oriented, Graph, Key-Value, Wide-Column, Free and Proprietary) were considered for clustering.

Table 4: Cluster Composition.

| Class I | Class II | Class III | Class IV | Class V | Class VI |
|---|---|---|---|---|---|
| Aerospike [67] | AllegroGraph [65] | Amazon Neptune [68] | ArangoDB [70] | Accumulo [66] | BerkeleyDB [73] |
| Cassandra [77] | Cache [76] | AnzoGraph [69] | BaseX [72] | Clusterpoint Database [80] | BigTable [75] |
| CDB or Constant Database [78] | Cloudant [79] | Azure Tables [71] | CouchDB [83] | CouchBase Server [82] | GridGain Systems [93] |
| etcd [90] | Coherence [81] | DataStax Enterprise Graph [86] | CrateIO [84] | HBase [4] | NoSQLz [117] |
| FoundationDB [92] | CosmosDB [85] | Dynamo [88] | ElasticSearch [89] | HyperTable [98] | OpenLink Virtuoso [119] |
| GT.M [54] | DocumentDB [87] | HyperGraphDB [97] | eXist [91] | MongoDB [6] | |
| Hibari [96] | IBM Informix [99] | HyperGraphDB [97] | Jackrabbit [104] | RethinkDB [128] | |
| IBM Informix C-ISAM [100] | Lotus Domino [110] | InfiniteGraph [102] | OrientDB [122] | TokuMX [141] | |
| Ignite [101] | Marklogic [111] | JanusGraph [105] | PostgreSQL [123] | TerraStore [143] | |
| InfinityDB [103] | Microsoft SQL Server [114] | KAI [106] | Sedna [135] | | |
| LevelDB [108] | MUMP Database [115] | Neo4j [116] | Solr [137] | | |
| Lightening Memory-Mapped Database (LMDB) [109] | ObjectDatabase++ [118] | Oracle Spatial and Graph [121] | | | |
| Memcached [112] | Oracle NoSQL Database [120] | Sparksee [138] | | | |
| MemcacheDB [113] | Qizx [125] | Sqrrl [139] | | | |
| Project Voldemort [124] | RavenDB [126] | XAP [146] | | | |
| Redis [127] | RocketU2 [130] | | | | |
| Riak [129] | SAP HANA [132] | | | | |
| RocksDB [131] | SimpleDB [136] | | | | |
| Scalaris [133] | | | | | |
| ScyllaDB [134] | | | | | |
| Tarantool [140] | | | | | |
| Tokyo Cabinet and Kyoto Cabinet [142] | | | | | |

Cluster counts for the four configurations presented in Table 3 namely, All, DMCAP, CAPFP and DMFP, were analyzed to infer that the same for 'All' were more evenly distributed as compared to others. This re-validates the fact that all the features must be considered for proposing a good classification scheme. Analysis of data shows that the count of Cluster 0 for 'All', which is the largest among all the other clusters decreases from n=3 to n=6 and then again, increases for n=7. Moreover, for n=7, the counts for Cluster 3, 4 and 5 remain the same. Therefore, n = 6 is taken as the best value for number of clusters. The created clusters are named classes and are shown in Table 4. Analysis of the cluster components reveals that the key features of each of the cluster are as follows:

- Class I contains NoSQL solutions that support key-value or wide-column data models and are proprietary.
- Class II contains NoSQL solutions that do not support wide-column data model and are free.
- Class III contains NoSQL solutions that do not support document-oriented data model and ensure partition tolerance.
- Class IV contains NoSQL solutions that support document-oriented data model, ensure availability and are proprietary.
- Class V contains NoSQL solutions that do not support graph data model, ensure consistency and partition tolerance and are proprietary.
- Class VI contains NoSQL solutions that ensure consistency and are free.

### 7.0 Choosing a NoSQL Solution for a Big Data System

As mentioned previously, for the sake of analysis, applications have been divided into 7 categories out of which 6 categories are specific application domains while the last category includes all unclassified works. All of the reviewed application papers used one or more of the listed NoSQL solutions. The base dataset was correspondingly appended with a column name 'Application Supported'. If the application type is supported by the NoSQL solution, then the data value of 'Application Supported' column for the cell corresponding to NoSQL solution was set to 1. On the other hand, if existing literature does not support the use of a particular NoSQL solution for an application category, then the corresponding value is set to 0. Such datasets were made for all the categories of applications.

Random Forest Classification [299] was performed for 6 application-specific datasets to compute feature importance of the 9 features with respect to six categories of applications namely smart cities, social network analysis, geospatial applications, life sciences research, healthcare and business intelligence. For all the experiments, the data was split into 75% and 25% for training and testing, respectively. In the Random Forest Classification performed for the dataset, Gini Importance [301] was used for determination of feature importance. This measure calculates the ratio of tree splits in which the feature is included and sample count for the split. The sum of these values is calculated for all the trees created in Random Forest to compute the final value of feature importance. Bar charts representing relative feature importance of the 9 features for each of these application areas are given in Fig. 5.

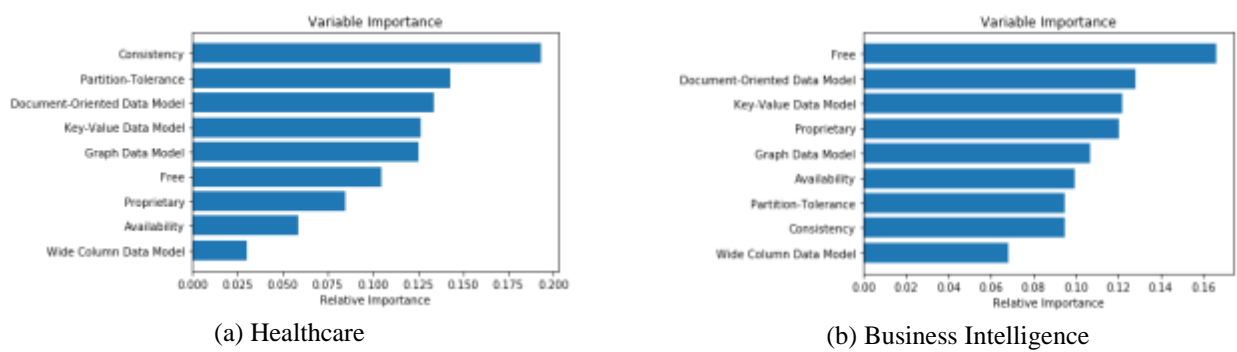

(a) Healthcare  (b) Business Intelligence

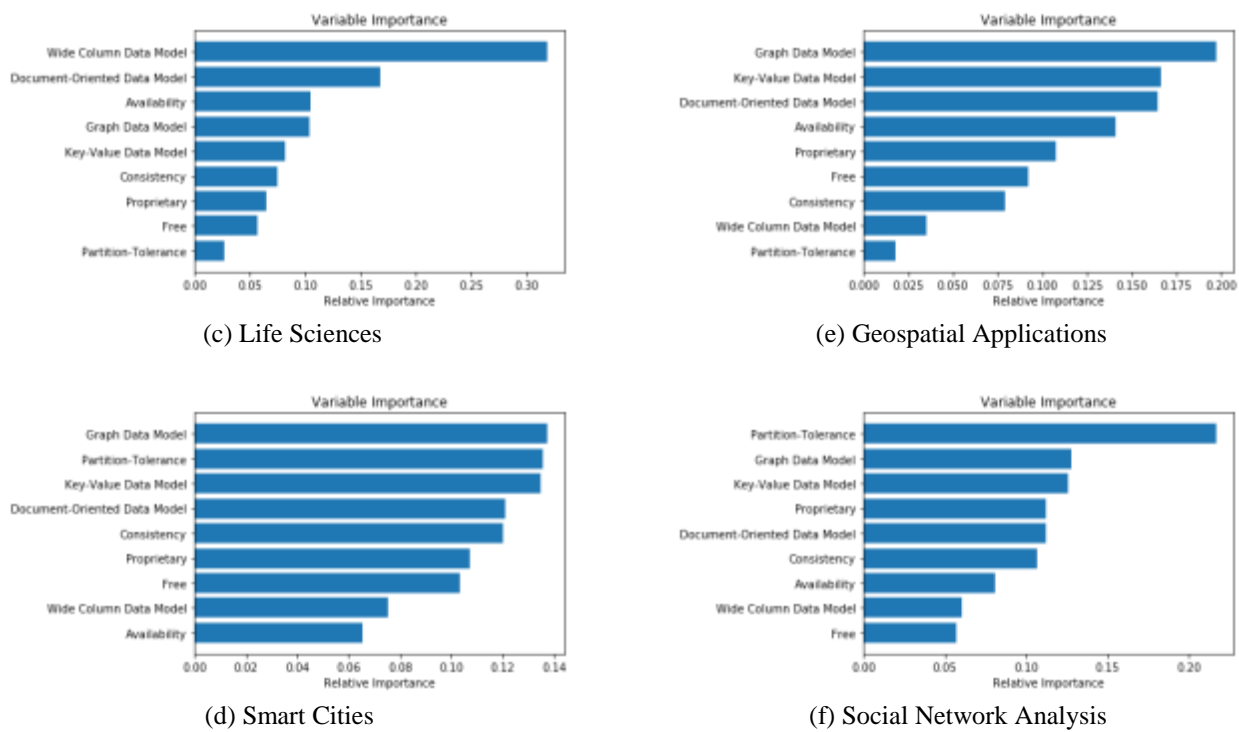

Fig. 5. Feature Importance for Different Application Areas

It is important to mention that feature importance values for all features are comparable for business intelligence (BI) applications, except for 'Free', which can be explained by the fact that budget is one of most crucial requirements in a BI application. Moreover, all the features show comparable values for smart cities. Theoretically, comparable feature importance values can be linked to the fact that business intelligence entails organizational as well as business-specific solutions. Similarly, smart cities also include a range of applications from waste management to smart manufacturing. Therefore, solution design can be of varied types and interests. Moreover, consistency, wide-column data model, partition tolerance and document-oriented data model are identified as the most relevant feature for healthcare, life sciences, social network analysis and geospatial applications, respectively.

Decision tree classification [155] was used as classification model for prediction of a NoSQL solution's suitability for an application area. Fig. 6 shows a visualization of the decision trees for respective applications. The blocks are symbolic of testing if a feature is present or absent. The labels 'Present' and 'Not Present' are indicative of whether the solution being tested has the feature specified in the preceding block. 'Suitable' and 'Not Suitable' are used as classes that indicate if the NoSQL solution being tested is found suitable or not suitable for the application area concerned. The accuracy obtained for decision tree classification outperformed the same for random forest classification, which means that the created trees for random forest had lesser correlation between them. Therefore, it has been established that decision tree classification is an appropriate prediction model for determining suitability of a NoSQL solution for a specific application area. The accuracy value comparison is provided in Table 5 and illustrated in Fig. 7.

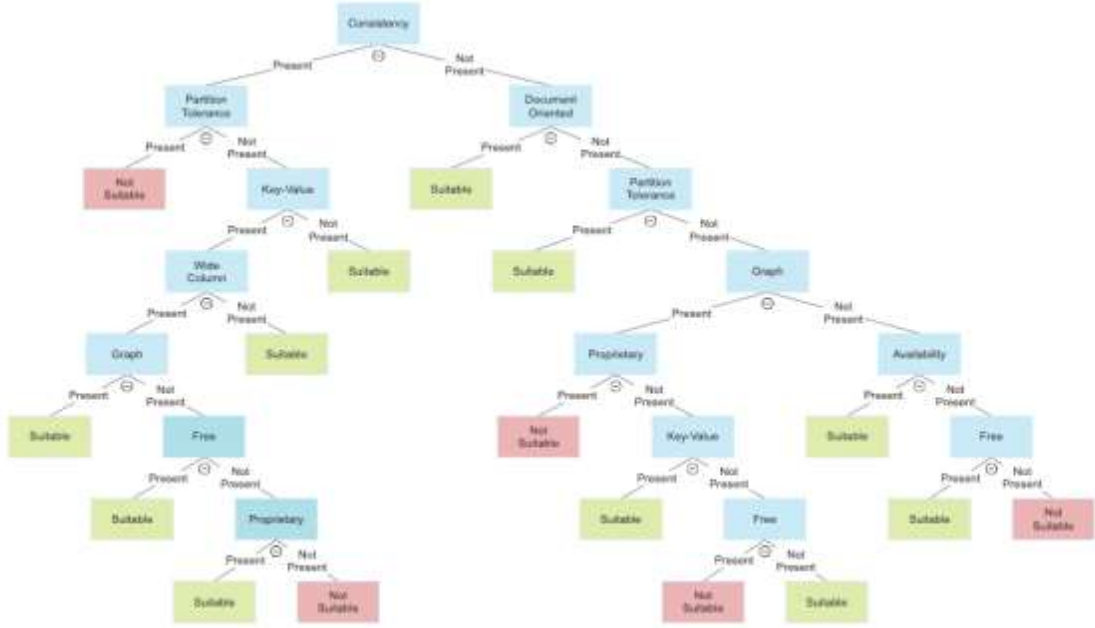

(a) Healthcare

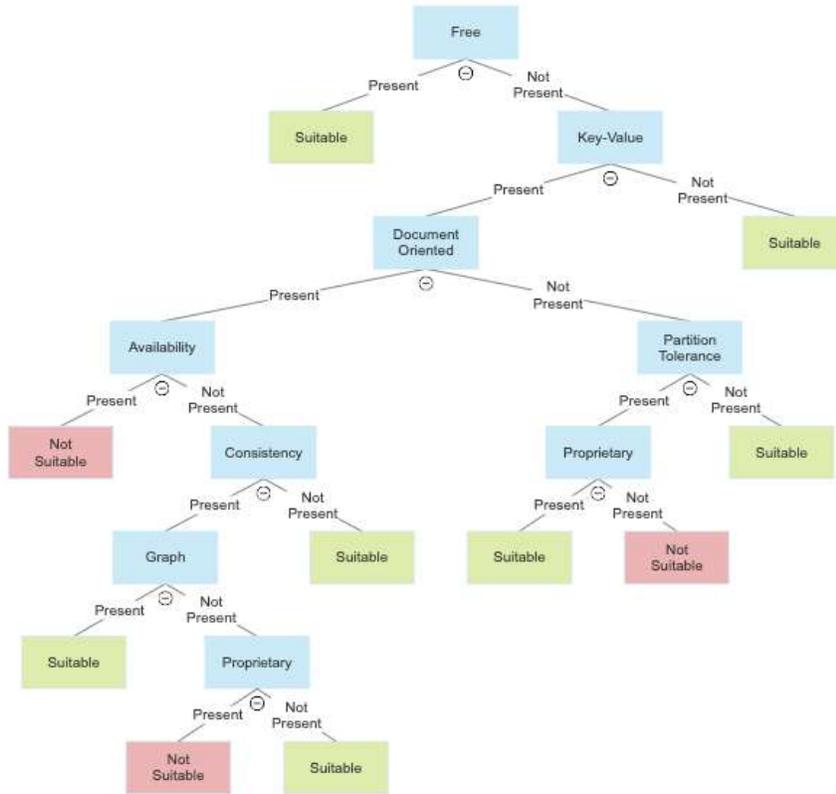

(b) Business Intelligence

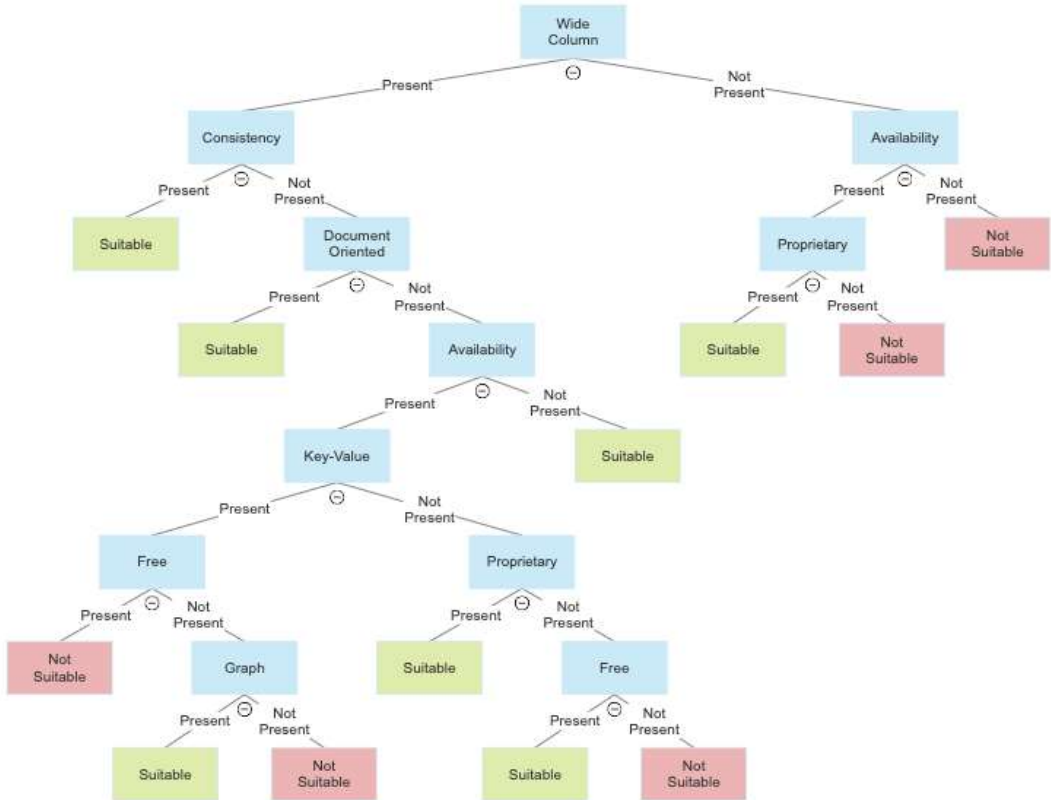

(c) Life Sciences

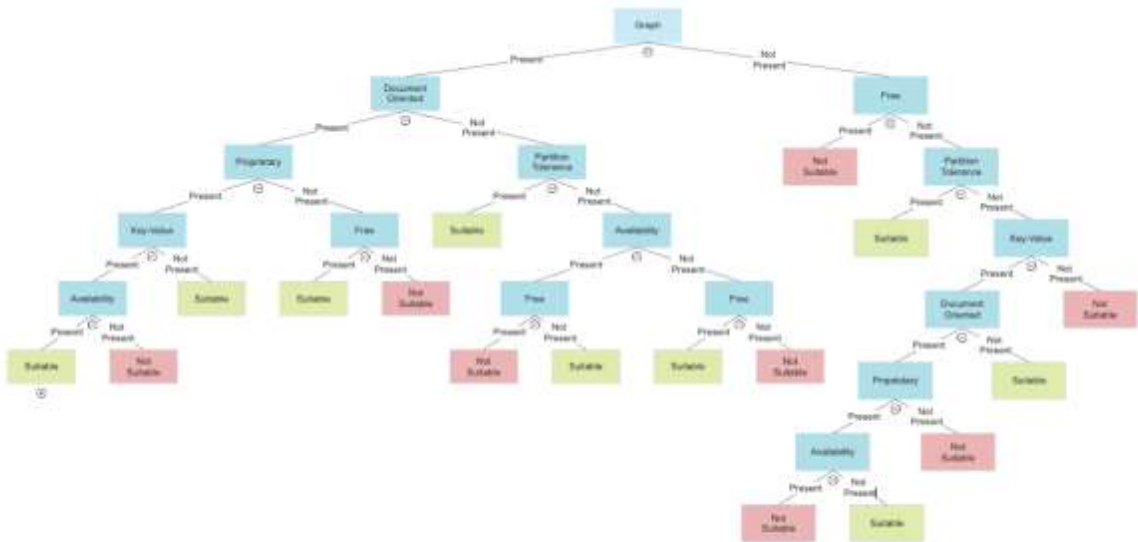

(d) Smart Cities

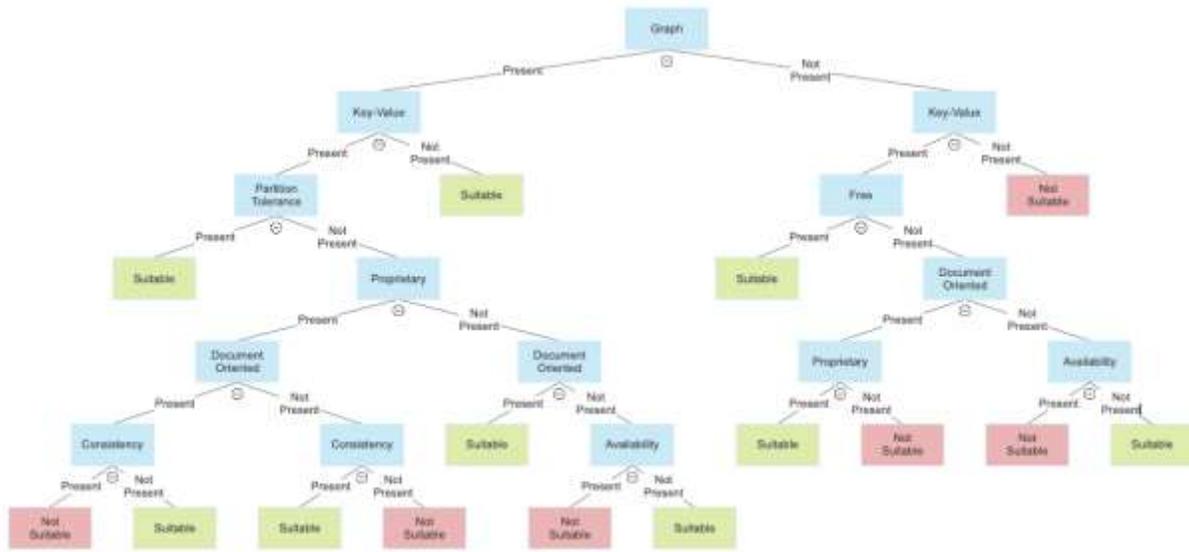

(e) Geospatial Applications

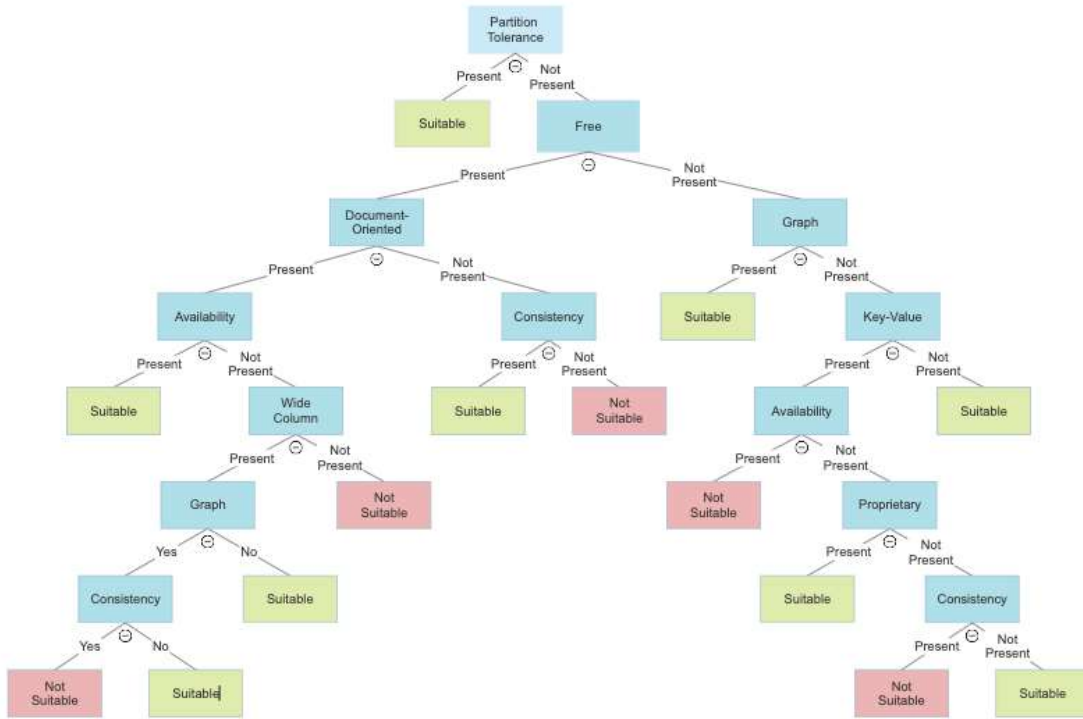

(f) Social Network Analysis

Fig. 6. Decision Trees for Different Application Areas

Table 5: Decision Tree Classification vs. Random Forest Classification

| Application | Random Forest | Decision Tree |
|---|---|---|
| Business Intelligence | 81.66% | 86.67% |
| Geospatial Applications | 88.33% | 93.33% |
| Healthcare | 75% | 88.33% |
| Life Sciences | 85% | 86.67% |
| Smart Cities | 65% | 78.33% |
| Social Network Analysis | 65% | 80% |

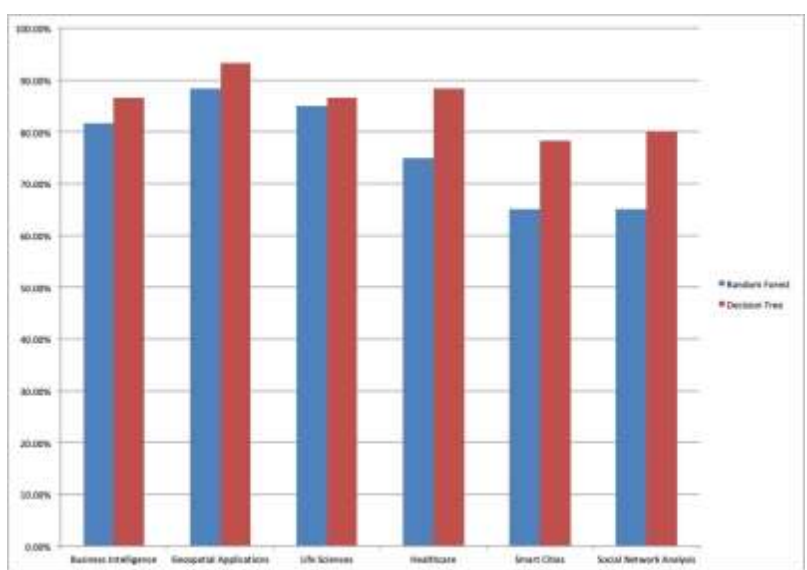

Fig. 7. Comparison of Accuracy for Decision Tree and Random Forest Classifications

**8.0 Discussion**

Design and development of a big data application that can resolve real world problems and prove to be a viable solution is dependent on the base technologies chosen for the creation of a heterogeneous storage and computing environment. This paper's scope is limited to comparison and analysis of NoSQL solutions available for big data systems. In view of specific storage challenges posed by big data like scalability, availability, integration and security, traditional systems are deemed incapable to handle the existing data scenario. NoSQL has proven to be a viable solution in this regard. Few of the significant features of NoSQL that prove the feasibility of its usage for big data storage and management include aspects like easily scalable systems, flexible data modeling, high availability and provisioning of required performance considering the fact that most modern systems handle static as well as real-time data. Features of NoSQL like dynamic schema, auto-sharding, automatic replication and integrated caching abilities mitigate the challenges posed by big data to traditional systems.

Understandably, data is the heart of the system and modeling data is the most important design activity for optimum system performance and functionality. In view of this, one of the most crucial technological decisions to be made while designing a big data application includes selecting a NoSQL solution. This decision requires determination of the data model of the application. The right data model for an application depends on the structure and composition of application data. The success of a big data application can be greatly impacted by a mismatch in the data model of the application and that of the chosen NoSQL solution. For instance, if the application's data can be represented in the form of a graph, then the graph model is most appropriate data model for the application. Each data model best suits to a specific set of applications and requirements.

The CAP theorem also suffers from several shortcomings because of the simplistic definitions of consistency, availability and partition-tolerance. For instance, transactions involving multiple objects are not dealt with when considering consistency [300]. Moreover, only partition tolerance is considered while other kinds of faults might also occur in the system and there is no contemplation on latency. Table 2 shows that there are some exceptions to the CAP theorem. While solutions like Solr provide only availability, RavenDB, MarkLogic, FoundationDB and Ignite provide all the three characteristics. There also exist some solutions like CosmosDB that provide variable and configurable consistency. This brings us to the conclusion that the existing classification schemes are insufficient when considered independently. Considering this, a novel classification scheme that uses identified features of the concerned NoSQL solution is proposed in this paper.

The market is flooded with solutions and technologies that provision a combination of customizable storage, acquisition, processing and visualization [188] facilities to the developer. This decision is based on many factors that can be technical and non-technical in nature. Once the right data model is determined for a big data problem, a solution that supports the data model along with the desired features as per the requirements of the big data system need to be found. An application may have specific requirements with respect to features like scalability and security. Choosing the right distribution model becomes an important consideration for an application that requires scalability as a fundamental requirement.

Scaling of read operations is supported by master-slave architecture. However, if scaling of both read and write operations is desired, then peer-to-peer architecture is a better option. Some NoSQL solutions can scale well like Cassandra whereas others may be memory-based and fail to scale across machines. Moreover, the use of NoSQL databases also has some security issues that must be considered and mitigated before a solution can be developed and deployed using the same. Therefore, adding more features to the proposed classification scheme and prediction model can be useful in making the model generic and this is proposed as future work. A decision-tree based prediction model is proposed in this paper for determining the suitability of NoSQL solution for an application area on the basis of the features it supports.

This work suffers from some limitations. Firstly, possible applications are grouped at a very coarse granularity while applications within these categories may be very different. Smart city and health have a variety of data processing needs including data privacy concerns, streaming data ingestion and complex analytic query processing requirements. For instance, there are healthcare and smart city applications which uses network related data and require graph data model. There are also applications in healthcare and agriculture that requires real-time or search analytics storage needs. In order to accommodate these, the comparative analysis can be further performed for lower level details such as read-write efficiency and storage structure. Some of the low level features such as sharding and caching have been discussed briefly but not included in the feature set.

Besides this, the generation of dataset is purely based on literature survey, which may raise some data quality concerns. Moreover, most application areas now deploy hybrid storage systems and have a variety of computational aspects, which require different types of storage systems. These concerns can be addressed by improving the data generation process. A field study based on some industry use cases can be used for this purpose. As NoSQL database systems and their applications evolve rapidly, the results of this paper were published at http://www.p-nasa.com/, and any updates on dataset and prediction model will be regularly updated on the website.

In addition, technological decisions are also driven by technical expertise. It has been found that developers accustomed of working on a technology are expected to choose it over solutions that might provide better performance. However, owing to project requirements, they might have to switch to better solution, but this wastes development time and effort. Efforts must be made to alleviate such issues.

## 9.0 Conclusion and Future Work

Review of existing literature suggests that both data model-based classification and CAP theorem are insufficient to provide discrete classification criteria for NoSQL solutions. It is proposed that hybrid feature categories must be used for discrete classification of solutions to make the decision-making process simpler for use-cases that include multi-dimensional data, which might require multiple base data models to design. In line with this, cluster analysis of 9 features (document-oriented data model, graph data model, key-value data model, wide-column data model, consistency, availability, partition tolerance, free and proprietary) of identified NoSQL solutions is used to create a unique classification scheme. Cluster analysis is used as the basis for creation of cluster categories.

This paper provides feature analysis of 80 NoSQL solutions to facilitate decision making in this regard. Moreover, this work also analyzes applications of individual NoSQL solutions and presents a prediction model that can be used to predict if a NoSQL solution is appropriate for a class of applications. The results are published at http://www.p-nasa.com. The website is in its preliminary phase of development and shall be improved in aesthetics and functionality in due course of time. The accuracy of the decision tree classification-based prediction model is more than 78% for all application areas. However, efforts to improve the accuracy of the prediction model shall be made in the future. Besides this, benchmarking of these solutions and their performance analysis for a complete quantitative comparison can be performed as future work. Identifying more features and adding more NoSQL solutions to the dataset can enhance the classification scheme and prediction model. This shall improve the quality and size of dataset used for development of prediction model.

**Acknowledgements**

This work was supported by a grant from "Young Faculty Research Fellowship" under Visvesvaraya PhD Scheme for Electronics and IT, Department of Electronics & Information Technology (DeitY), Ministry of Communications & IT, Government of India.

MongoDB and PostgreSQL for spatio-temporal data.